\documentclass[twocolumn]{aastex63}
\usepackage{floatrow}%
\usepackage{lipsum}
\usepackage{mwe}
\usepackage{color}
\usepackage[caption=false]{subfig}

\received{October 30, 2020}
\revised{February 22, 2021}
\accepted{\today}
\submitjournal{AJ}

\shorttitle{ EDGES Low-Band Beam Model}
\shortauthors{Mahesh et al.}

\begin{document}

\title{Validation of EDGES Low-Band Antenna Beam Model}

\correspondingauthor{Nivedita Mahesh}
\email{nivedita.mahesh@asu.edu}

\author[0000-0002-0786-7307]{Nivedita Mahesh}
\affiliation{School of Earth and Space exploration, Arizona State University, Tempe, AZ 85287, USA}

\author{Judd D. Bowman}
\affiliation{School of Earth and Space exploration, Arizona State University, Tempe, AZ 85287, USA}

\author{Thomas J. Mozdzen}
\affiliation{School of Earth and Space exploration, Arizona State University, Tempe, AZ 85287, USA}

\author{Alan E. E. Rogers}
\affiliation{Haystack Observatory, Massachusetts Institute of Technology,Westford, Massachusetts 01886, USA}

\author{Raul A. Monsalve}
\affiliation{Department of Physics and McGill Space Institute, McGill University, Montreal, QC H3A 2T8, Canada}
\affiliation{School of Earth and Space exploration, Arizona State University, Tempe, AZ 85287, USA}
\affiliation{Facultad de Ingeniería, Universidad Católica de la Santísima Concepción, Alonso de Ribera 2850, Concepción,Chile}


\author{Steven G. Murray}
\affiliation{School of Earth and Space exploration, Arizona State University, Tempe, AZ 85287, USA}

\author{David Lewis}
\affiliation{School of Earth and Space exploration, Arizona State University, Tempe, AZ 85287, USA}



\begin{abstract}


The response of the antenna is a source of uncertainty in measurements with the Experiment to Detect the Global EoR Signature (EDGES).  We aim to validate the electromagnetic beam model of the low-band (50-100~MHz) dipole antenna with comparisons between models and against data. We find that simulations of a simplified model of the antenna over an infinite perfectly conducting ground plane are, with one exception, robust to changes of numerical electromagnetic solver code or algorithm.  For simulations of the antenna with the actual finite ground plane and realistic soil properties, we find that two out of three numerical solvers agree well.  Applying our analysis pipeline to a simulated driftscan observation from an early EDGES low-band instrument that had a 10~m~$\times$~10~m ground plane, we find residual levels after fitting and removing a five-term foreground model to data binned in Local Sidereal Time (LST) average about 250~mK with $\pm$40~mK variation between numerical solvers.  A similar analysis of the primary 30~m~$\times$~30~m sawtooth ground plane reduced the LST-averaged residuals to about 90~mK with $\pm$10~mK between the two viable solvers. More broadly we show that larger ground planes generally perform better than smaller ground planes. Simulated data have a power which is within 4$\%$ of real observations, a limitation of net accuracy of the sky and beam models. We observe that residual spectral structures after foreground model fits match qualitatively between simulated data and observations, suggesting that the frequency dependence of the beam is reasonably represented by the models. We find that soil conductivity of 0.02~Sm$^{-1}$ and relative permittivity of 3.5 yield good agreement between simulated spectra and observations.  This is consistent with the soil properties reported by \citet{Sutinjo_2015} for the Murchison Radio-astronomy Observatory, where EDGES is located.
 

\end{abstract}



\section{Introduction} 
\label{sec:intro}


Observations of the early Universe spanning the first billion years, from Cosmic Dawn, when the first stars formed, to the Epoch of Reionization (EoR), when the collective radiation from galaxies influenced the entire universe, can shed light on the formation of the first stars and the thermal and ionization history of the intergalactic medium (IGM). Much useful information, such as the growth of structure, baryon densities, heating and ionization of baryons, and the physical properties of the first light sources, can be gained by studying these periods.  

There are many ongoing astrophysical probes of these epochs ranging from direct to indirect techniques including Gunn-Peterson trough observations towards high-redshift quasars \citep{Fan_2006, 10.1093/mnras/stu2449}; tracking the evolution in the luminosity function of Lyman-$\alpha$ galaxies \citep{Zheng_2017}; detection of the EoR signature in the Cosmic Microwave Background (CMB) anisotropies \citep{2016A&A...594A..13P, 2013ApJS..208...19H}; and upper limits on the "patchy" kinematic Sunyaev-Zeldovich effect \citep{2016A&A...594A..13P,Zahn_2012}. The redshifted 21~cm radio signal due to the spin flip transition of the neutral hydrogen atom is of particular interest because it is a direct probe of the IGM in the EoR. The spatial and temporal fluctuations in the 21~cm signal that traces the cosmological evolution are caused by Wouthuysen-Field \citep{1952AJ.....57R..31W, 1958PIRE...46..240F} coupling of the spin to the kinetic temperature via Lyman-$\alpha$ photons, gas heating via X-rays, and reionization via ultra-violet radiation. 

There are two approaches to measuring the brightness temperature of the redshifted 21~cm signal.  One approach is to statistically measure the spatial and spectral fluctuations of 21~cm emission and absorption using interferometers. Several experiments have pioneered this approach, including the LOw Frequency ARray (LOFAR; \citealt{van_Haarlem_2013}), the Giant Meterwave Radio Telescope (GMRT; \citealt{Paciga_2011}), the Hydrogen Epoch of reionization Array (HERA; \citealt{DeBoer_2017}), the Murchison Widefield Array (MWA; \citealt{2013PASA...30....7T,Bowman_2013}), the 21 Centimeter Array (21CMA; \citealt{Zheng_2016}), the Owens Valley Radio Observatory-Low Wavelength Array(OVRO-LWA; \citealt{Eastwood_2019}), and the now decomissioned Precision Array to Probe the Epoch of Re-ionization (PAPER;\citealt{2015ApJ...809...61A, 2015ApJ...809...62P}).

The second approach consists of using a single antenna element to measure the sky-averaged or ``global''  brightness temperature of the cosmological 21~cm signal. This method is pursued by the Experiment to Detect the Global EoR Signature (EDGES) \citep{bowman2018absorption, bowman2010lower, 2008ApJ...676....1B}, the Shaped Antenna measurement of background Radio Spectrum (SARAS) \citep{2013ExA....36..319P, 2017ApJ...845L..12S}, the Probing Radio Intensity at high-Z from Marion (PRI$^Z$M) \citep{2019JAI.....850004P}, the Large-aperture Experiment to detect the Dark Age (LEDA) \citep{Bernardi_2016}, the Cosmic Twilight Polarimeter (CTP) \citep{2018AAS...23210204N}, and the now decommissioned Broadband Instrument for Global Hydrogen Reionisation Signal (BIGHORNS) \citep{2015PASA...32....4S}.

Detection of the global 21~cm signal has been limited by the interaction of the instrument with strong foreground emission. The 21~cm signal is broadband below $200$~MHz and is expected to have a peak absolute amplitude between 10 and 200~mK, whereas Galactic and extragalactic continuum foregrounds from synchrotron and free-free emission are approximately four orders of magnitude larger, with typical sky temperatures of 2000~K at 75~MHz away from the Galactic center. The foregrounds generally exhibit smooth, power-law-like spectra that must be subtracted from observations to reveal the 21~cm signal. For both interferometric and global experiments, antenna beam variation with frequency, or "chromaticity", can couple angular variations in foregrounds into spectral structures, making the foregrounds more covariant with the 21~cm signal and, therefore, more difficult to separate.

In \citet{bowman2018absorption}, we reported evidence for detection of a 21~cm global signal from the era of Cosmic Dawn using the EDGES low-band instruments. However, the absorption feature seen by EDGES was narrower and sharper than predicted and its amplitude is 2-3 times larger than standard astrophysical models.  We considered possible alternative explanations for the observed feature, such as an instrumental artifact caused by receiver calibration errors or unmodeled frequency-dependence of the antenna beam converting angular foreground power into spectral features. 
Others have also proposed alternative instrumental or analysis explanations \citep{Bradley_2019, sims, Hills_2018} or polarized foregrounds \citep{10.1093/mnras/stz2425}.  
We conducted a number of validation tests that disfavoured instrumentation effects, including repeating the measurement with multiple instruments, with the antenna at different rotation angles, and using multiple calibration solutions.  We disfavoured unmodeled antenna beam chromaticity by extracting the absorption feature at different LSTs when the angular structure and total power of the Galactic foregrounds were different.  The EDGES high-band instrument, which uses a scaled copy of the low-band antenna, has observed at 90-190~MHz.  These observations do not show a comparable structure at the scaled frequencies.  They have been used to further constrain parameters of early galaxies, including UV ionizing efficiency, halo virial temperature, and soft X-ray luminosity \citep{Monsalvetwo,2018ApJ...863...11M,2019ApJ...875...67M}.

The antenna beam remains a source of potential uncertainty and error in EDGES data analysis. In order to provide estimates of the intrinsic brightness temperature of the sky, EDGES observations are corrected for the effects of the frequency-dependent antenna beam sweeping across the angular structure of the foregrounds.  We use an antenna beam model and sky model to estimate these effects \citep{Mozdzen_2016,Mozdzen_2019}. Removing this contribution from the data can introduce errors due to insufficient knowledge of the beam and sky models.  We minimize errors associated with the beam by using a simple, planar dipole antenna design for all EDGES instruments, called a ``blade'' dipole \citep{oliner2207}.  This antenna has two advantages: it is relatively easy to simulate and, due to its compact size and simple design, its properties should change slowly and smoothly with frequency.

Motivated by the possible effects of frequency dependent chromatic antenna beam structure on the observed spectrum, and to facilitate potential improvements to the instrument design, we validate our beam models to gain better knowledge of the EDGES antenna beams and how they change with frequency. We show in Section~\ref{best-case_fom} that for the level needed in 21~cm cosmology, the chromaticity of the beam after removing a three-term polynomial must be at the level of 0.05 (linear units), or approximately 0.01-0.1~dB across the primary response region of a dipole antenna. 

There are currently no techniques available that can provide emperical measurements of a low-frequency antenna beam at the needed level.  Anechoic chamber measurements typically have an accuracy of order 0.3~dB along the boresight, reducing to 0.8~dB or worse in the sidelobes \citep{8778885}.  This is not sufficient to characterize the fine features in the beam as needed for 21~cm cosmology. Further limiting the use of anechoic chamber measurements is that the local environment around low-frequency antennas influences their properties, necessitating in-situ measurements.  However, in-situ measurements are very challenging. The SARAS~2 global 21~cm experiment has made in-situ measurements of its antenna gain, achieving only about 10$\%$ accuracy \citep{Singh_2018}.  The EDGES antennas are not embedded in an array or near a large single-dish telescope, limiting the opportunities to use point-source holography \citep{2016SPIE.9906E..0DB} and driftscan techniques \citep{2012AJ....143...53P} that rely on isolating individual point sources as they drift overhead in order to create a beam map.  Transmissions from the ORBCOMM constellation of low-Earth orbit satellites have been used to successfully map antenna beams for the MWA \citep{2015RaSc...50..614N, 2018PASA...35...45L}, but at 137~MHz, the transmitter frequency is above the EDGES low-band range and, at only one frequency, could not provide full frequency-dependent mapping of the beam. Drone-mounted transmitters are under development to overcome the limitations of the ORBCOMM technique and show much promise \citep{2017PASP..129c5002J, 2015PASP..127.1131C, 2015ExA....39..405P, 2014IAWPL..13..169V}, but they have not yet demonstrated the required accuracy.  Therefore, to better understand the EDGES low-band beam, we must rely on numerical simulations for the time being.   

In this paper, we simulate the EDGES blade dipole antenna, used in the low-band instruments, with three different electromagnetic (EM) solvers and commercial software packages to assess the accuracy and repeatability of the simulated antenna response. In particular, the EM solvers we use are the Method of Moments, the Finite Element Method, and the Finite Difference Time Domain method. We quantify the chromatic structure in the beam and compare it with the variation seen in observations. We begin in Section 2 with an overview of the EDGES low-band system, describe the EM Solvers we use to obtain the beam solutions and introduce the metrics to quantify the chromaticity of the beam models. In Section~3, we present the results of modeling the antenna over an ideal ground plane with the three EM techniques. We also discuss the refinement strategies implemented to get consistent results across the different EM techniques. We use the metrics introduced in Section~2 to compare all of the beam solutions.
Section~4 shows the antenna gain results and the associated chromaticity when modeling with a finite ground plane and realistic soil beneath. The results for the chromaticity from simulations are compared with the actual data collected using one of the low-band instruments. In this section, we also confirm the electrical properties of the soil at the EDGES site by analyzing the agreement between simulated and measured sky spectra. In Section~5, we present our discussions and conclusions.

\section {Methods}

We use several EM solvers to obtain the response of the EDGES low-band antenna.  Our goal is to calculate and analyze the chromaticity of the antenna beam pattern by obtaining beam solutions as a function of frequency. The beam pattern is used in the data processing pipeline on the obtained sky spectra. Hence it is essential to reduce the uncertainties in the beam model that could lead to errors in data analysis. We carry out a number of simulations to cross-validate the solutions from different packages and different solvers within each package.   

\subsection{Electromagnetic Solvers}

A variety of numerical methods exist for the analysis of electromagnetic fields associated with an antenna. They are based on the solution of Maxwell's equations or equations derived from them. Maxwell's equations are fundamental equations for electromagnetic fields and they can be solved in integral and differential form \citep{godara2018handbook}. The different numerical methods can be classified based on the unknown quantity that they solve for.  The solution methods and solvers we use in this analysis are as follows:  

\textbf{Method of Moments (MoM)}.   The MoM solution solves for the field sources like currents and charges. These sources can be either physical sources, or mathematically equivalent sources introduced through various electromagnetic field theorems \citep{harrington1993}.  The electromagnetic fields, or the related potentials, are expressed in terms of these sources, usually through the Lorentz potentials. The expressions are integral forms, where the sources appear under some integrals, multiplied by appropriate functions, which are referred to as kernels. For example, the far-fields for the EDGES antenna over a perfect electric conductor (PEC) ground in vacuum are solved with a free-space Green's function kernel for the Lorentz potentials. The MoM technique is implemented in the FEKO\footnote{Altair FEKO - https://altairhyperworks.com/product/FEKO}, CST\footnote{Computer Simulation Technology - www.cst.com}-I (Integral solver), and HFSS\footnote{High Frequency Structure Simulator - https://www.ansys.com/products/electronics/ansys-hfss}-IE (Integral Equation) software packages.

\textbf{Finite Element Method (FEM)}.  Unlike the MoM technique, FEM solves directly for the electric or magnetic field vectors, or for quantities tightly related with them (e.g., the Lorentz potentials). The starting equations are Maxwell's equations in differential form or their derivatives (e.g., the wave equation) \citep{Zienkiewicz}. The unknowns are, hence, spread throughout the volume occupied by the fields. This volume is encompassed within a radiation box and extends up to the radiative region of the antenna i.e, the Fraunhofer distance.  The FEM technique is implemented in the HFSS and CST-F software packages. 

\textbf{Finite Difference Time Domain (FDTD)}. The FDTD method employs finite differences as approximations to both the spatial and temporal derivatives that appear in Maxwell’s equations (specifically Ampere’s and Faraday’s laws). The equations are solved in a cyclic manner: the electric field is solved at a given instant in time, then the magnetic field is solved at the next instant in time, and the process is repeated over and over again \citep{sullivan2013electromagnetic}. Since it is a time-domain method, solutions can cover a wide frequency range with a single simulation run (unlike the MoM and FEM techniques), provided the time step is small enough to satisfy the Nyquist–Shannon sampling theorem for the desired highest frequency. Like the FEM technique, the FDTD solves for the E- and H-fields directly. This method therefore uses a radiation box to compute the unknowns within the enclosed volume. Of the three packages evaluated, only CST-T supports this technique.  

The three techniques used across the three different commercially available software packages provide a total of six independent solutions to compare.  

\subsection{Antenna Model}
\label{sec_ant-model}
EDGES is located in the Murchison Radio-astronomy Observatory (MRO) in Western Australia. The experiment is based on a single-element broadband radio spectrometer design.  It uses multiple instruments to cover the frequency range of interest from 50 to 200~MHz.  The low-band instrument operates from approximately 50 to 100~MHz and the high-band operates from approximately 100 to 200~MHz. Recently, a mid-band instrument spanning an intermediate range of frequencies was added to the experiment.  

Each instrument uses a nearly identical copy of analog and digital electronics. All instruments use a scaled copy of the same blade antenna design that has been optimized for a particular subset of the frequency range.  The blade antenna is supported above a wire mesh ground plane by a fiberglass frame that has little impact on its electrical properties.  The blade dipole has low beam chromaticity according to our previous simulations \citep{Mozdzen_2015}.  The antenna has a maximum response toward zenith and its planar design reduces the horizon response compared to non-planar antennas. This reduces direct pick up of ground radiation and radio frequency interference. A Roberts transmission-line balun \citep{balun} is used to transition from the panels to the receiver. A capacitive top plate above the blade excitation point of the antenna is employed to improve the impedance match of the antenna to the receiver. At the bottom surrounding the Balun rods is a small rectangular metal shield that reduces the radiation from the vertical currents in the balun tubes, which are strongest at the base. 

\begin{figure}[ht!]
\centering
\includegraphics[width=\textwidth]{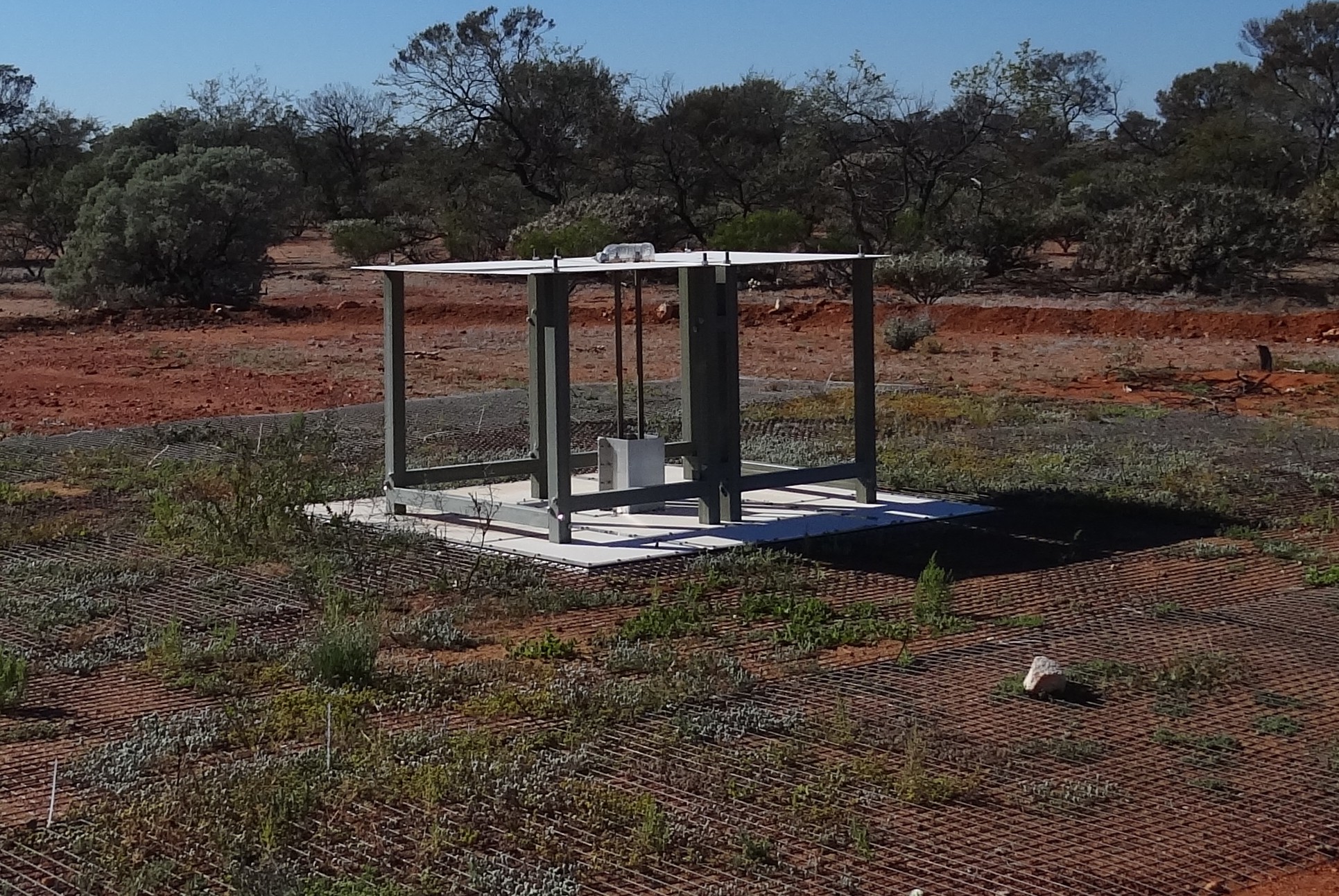}
\includegraphics[width=\textwidth]{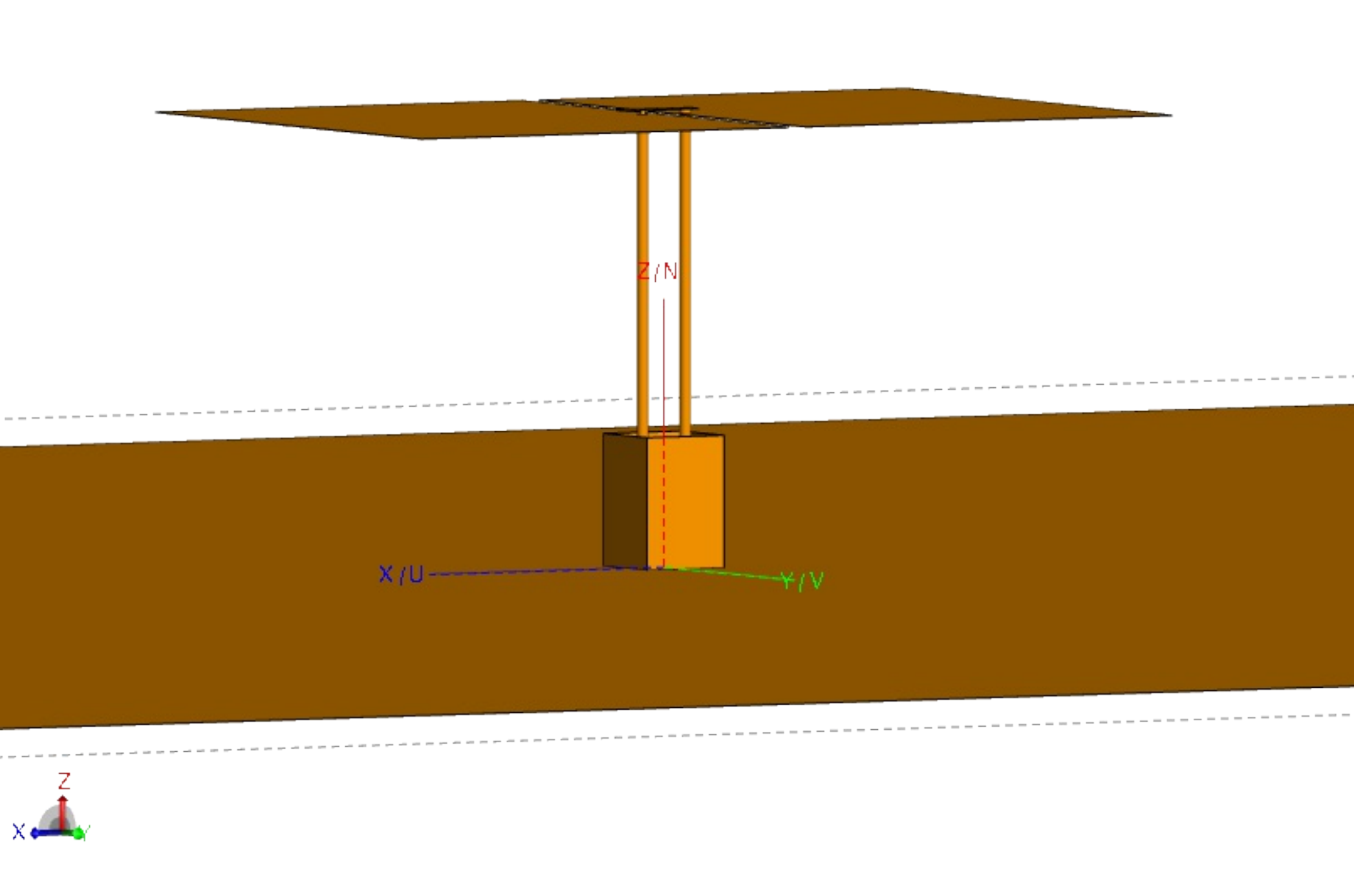}
\caption{[Top] EDGES low-band antenna located at the Murchison Radio-astronomy Observatory in Western Australia. The panels of the blade dipole have a balun at the center that collects the signals and directs them to the receiver, which is placed below the ground plane. The fiberglass tubes (not included in the EM model) visible in the picture are used as a frame to support the panels and the balun. The center portion of the ground plane, visible in white paint, is made of metal sheets and then connects to a galvanized steel wire mesh with 5~cm grid spacing. [Bottom] Model of the EDGES antenna used in FEKO based on the dimensions in Table~\ref{table-antenna-characteristics}.}
\label{Antenna-site}
\end{figure}

In this work, we will focus entirely on the EDGES low-band antenna.  A photo of the low-band antenna is shown in Figure~\ref{Antenna-site} and a summary of its  properties is presented in Table~\ref{table-antenna-characteristics}. 

\begin{table}[h!]
 \caption{Summary of EDGES low-band antenna characteristics}
 \centering
 \begin{tabular}{l c}
 \hline
  Parameter  & Value  \\
 \hline
   
   Panel length  &  125.2~cm  \\
   Panel width  & 96.4~cm   \\
   Gap between the panels & 4.4~cm\\
   Height (from the ground)  & 104~cm \\
   Original ground plane  & 10~m $\times$ 10~m   \\
   Extended ground plane & 30~m $\times$ 30~m \\
   Ground plane mesh grid size & 5~cm \\
   3~dB beam width @ 75MHz (E-plane; $\phi = 0^\circ$)  &  $72^\circ$\\
   3~dB beam width @ 75 MHz (H-plane; $\phi = 90^\circ$)  & $110^\circ$   \\
   Excitation azimuth & -7$^\circ$ \\
 \hline
 \label{table-antenna-characteristics}
 \end{tabular}
 \end{table}

The metal wire mesh ground plane is made from galvanized steel wires and has 5~cm of separation between mesh wires.  All wires are welded at the intersections with other wires for good conductivity.  The antenna and ground plane are aligned nearly north-south, with the excitation access of the antenna at an azimuth of $-7^\circ$.  Two different configurations of the wire mesh ground plane have been placed below the low-band dipole. The first deployment used a simple square of side 9.8~m. This version of the instrument acquired sky data from October 2015 through July 2016. The ground plane was then upgraded and extended to reduce the beam chromaticity of the antenna as discussed in Section~\ref{gp-imp}. The extended ground plane has a square mesh of side 20~m. Each side of the central mesh has four isosceles triangle extensions attached to it, evenly spaced along the edge. Each triangle extension has a base width of 4.8~m and height of 5~m. The extended ground plane has a total edge-to-edge extent of 30~m.  This version of the ground plane is used in the current low-band system and began data acquisition in September of 2016.

\begin{figure}[ht!]
\centering

\includegraphics[width=\textwidth]{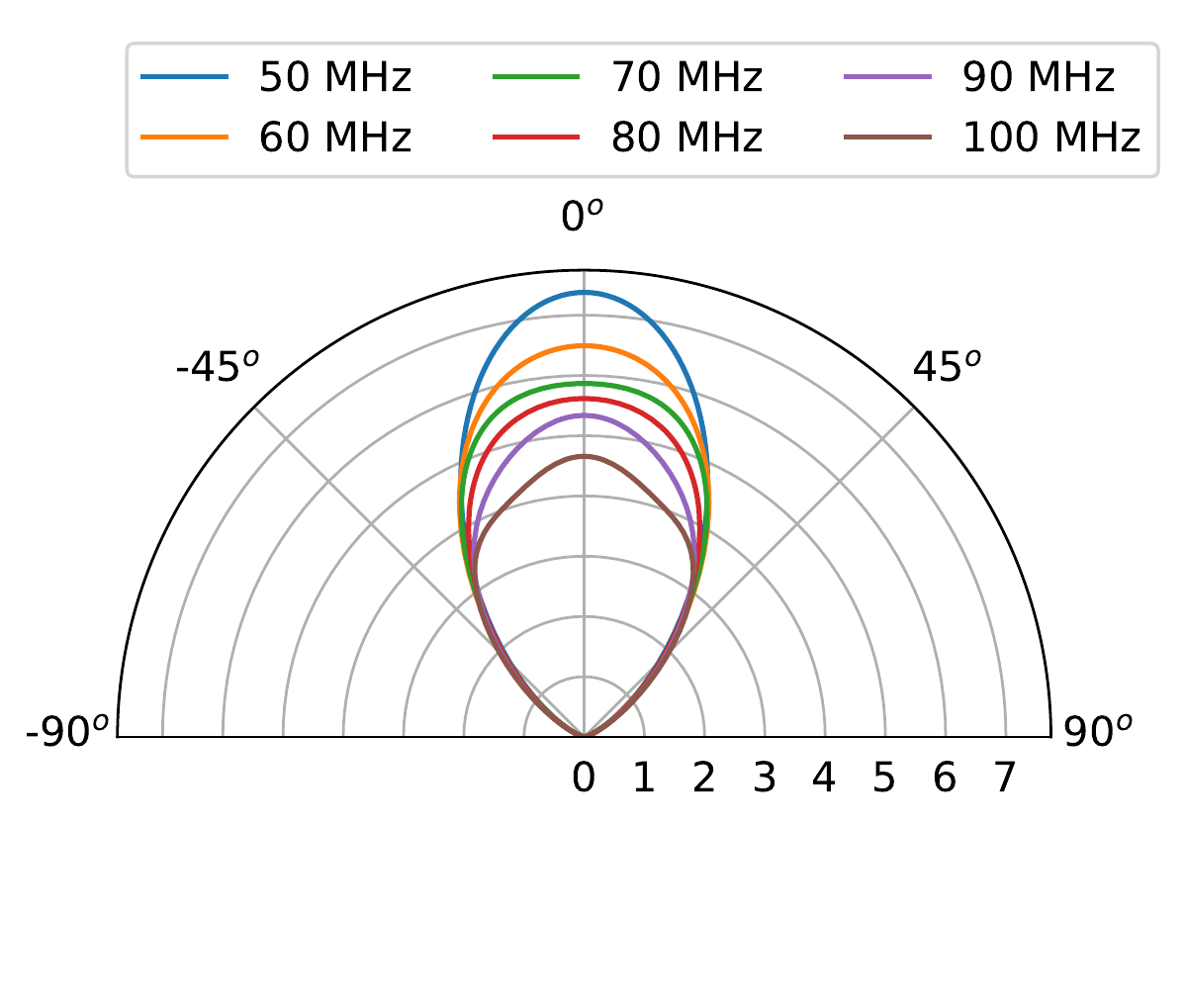}
\includegraphics[width=\textwidth]{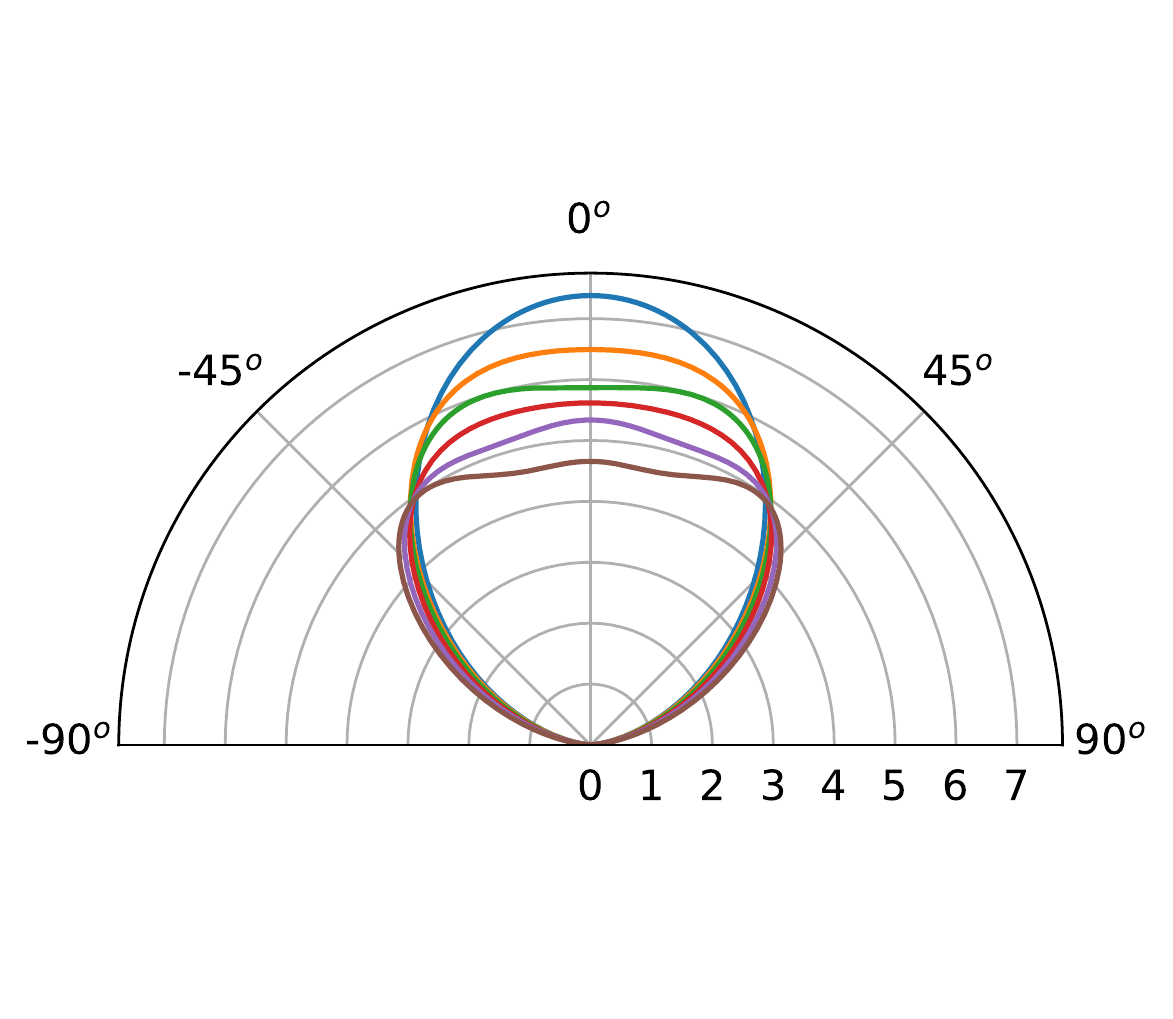}

\caption{ Gain as a function of elevation angle (theta) for two cuts through the beam at several frequencies.  The gain is plotted in linear scale. [Top] For cuts along the excitation axis of the dipole (E-Plane). [Bottom]  For cuts along the perpendicular axis (H-plane). These beam solutions were obtained by simulating the EDGES dipole over the 10~m~$\times$~10~m ground plane using FEKO.}
\label{polar-plot-f-var}
\end{figure}

In our EM models, we will neglect the fiberglass support structure and treat all parts of the antenna and balun as PEC. The gain and directivity of an antenna is related by the antenna/electrical efficiency factor ($\eta$), i.e, G=$\eta$D. So, for a lossless antenna ($\eta=1$) where the entire structure is modelled only with PEC as in our case, the gain and directivity will be the same. The antenna model is meshed using an adaptive meshing criteria at the highest simulated operating frequency (100~MHz) in each solver, which refines the mesh until the variation in the antenna reflection coefficient (S11) between refinements converges to less than 0.1~dB.  With this mesh setting, the solution for the far-field gain is calculated at 1$^\circ$~spacings in azimuth and elevation for frequencies between 40~and 100~MHz in 1~MHz steps. In Figure~\ref{polar-plot-f-var}, we show example cross-sections of the beam gain at several frequencies. Both E-plane (along the axis of excitation; $\phi=0^o$) and H-plane (perpendicular to the axis of excitation; $\phi=90^o$) cuts of the power gain of the beam are plotted. The variation of the curves with frequency indicates that the EDGES beam is chromatic and has to be modeled carefully. 

\bigskip

\subsection{Metrics and Figure of Merit}
\label{sec_fom}

We adopt two metrics to compare the beam solutions from the different solvers. These also quantify the beam variation with frequency. The first metric we use is to analyse the level of the residuals after removing a low-order polynomial fit to the gain as a function of frequency at specific azimuth-elevation viewing angles.  The second metric is the derivative of the beam gain along the frequency axis:
\begin{equation} 
\label{eqn_derivative}
\frac{G_{\theta, \phi}(\nu+\Delta \nu) - G_{\theta, \phi}(\nu)}{\Delta \nu},
\end{equation}
where $G_{\theta,\phi}(\nu)$ is the gain value of the beam at a given frequency ($\nu$) and, spherical coordinates ($\theta$, $\phi$), and $\Delta \nu$ is the small step in frequency.

We also introduce a Figure of Merit (FoM) that enables us to quantitatively compare the different beam solutions in connection with our science goals. This FoM uses simulated measurements of the antenna temperature as function of frequency obtained by convolving the antenna beam model with a sky model.   For the sky model, we use the \citet{1982A&AS...47....1H} all-sky brightness temperature map at 408~MHz and scale it to the desired frequency range using a spectral index of $\beta=2.55$, following:
\begin{equation}
    T_{sky}(\nu) = T_{408} \left( \frac{\nu}{\nu_{408}} \right)^{-\beta},
\end{equation}
where $T_{sky} (\nu)$ \footnote{We did not remove the contribution from the CMB to the map (2.725 K) during its frequency scaling because it is not significant for the results of this paper and because it is within the uncertainty range reported for the zero-level of the Haslam map.} is the temperature of the sky evaluated at frequency $\nu$ and $T_{408}$ is the sky temperature from the 408~MHz Haslam sky map. At each frequency for which we have evaluated the antenna beam patterns, we project the beam pattern onto the sky map using the MRO site latitude and a specific LST.  We integrate the product of the two over the sky coordinates to calculate the simulated antenna temperature at each frequency, according to:
\begin{equation} 
T_{ant}(\nu,n) = \frac{1}{4\pi}\int G(\nu,\theta,\phi) ~ T_{sky}(\nu,\theta,\phi,n) d\Omega,
\end{equation}
where $T_{{sky}}(\nu,\theta, \phi, n)$ is the brightness temperature of the sky at a given frequency, spherical coordinates and LST ($n$).  Finally,  we perform this computation in steps of 1$^\circ$ in LST to create a two-dimensional array of 61~frequency samples and 360~LST samples representing a simulated EDGES low-band (40-100~MHz) driftscan spanning 24~hours.  

Our FoM is the RMS of the residuals over the frequency range of 55 MHz to 97 MHz after fitting and removing a foreground model from each spectrum in the simulated driftscan data. The frequency range chosen to calculate the RMS excludes the band edges where the S11 of the blade antenna deteriorates quickly. The FoM (as a function of LST bin) is given by:  
\begin{equation}
\text{FoM(n)} =   \sqrt{\langle [T_{ant}(\nu,n) - T_{model}(\nu,n)]^2\rangle_v} 
\label{eqn-fom}
\end{equation}
and we use the foreground model,
\begin{equation}
T_{model}(\nu, n) = \left(\frac{\nu}{\nu_o}\right)^{-\beta}~\Sigma_{m=0}^{M-1} a_{m,n} \left[ \ln \left (\frac{\nu}{\nu_o} \right)\right] ^{m}
\label{linpoly}
\end{equation}
where $T_{model}(\nu,n)$ is the foreground model as a function of frequency and LST. The foreground model used is one of the models we often employ in our 21~cm signal searches. Thus, the FoM effectively captures the sensitivity of 21~cm signal searches to the antenna beam chromaticity. We refer to this foreground model, which is an approximation to the typical exponential log model used by theoretical analyses, as the LinLog model \citep{bowman2018absorption, memo_judd}. In this paper we use $M=5$ for equation \ref{linpoly}.The choice of a five-term polynomial in the FoM is motivated to yield a FoM that closely reflects the analysis that is performed to search for the global 21~cm signal. This helps us access the impact of the antenna beam on our science results.  The intrinsic sky noise spectrum is dominated by smooth synchrotron radiation. The foreground model absorbs the large-scale trends in the spectra, hence the FoM is not particularly sensitive to the choice of spectral index, $\beta$, used for the sky map scaling. 

We generate a FoM for the performance of the antenna pattern as a function of LST for each of the solvers.  We will also use an LST-averaged FoM that is found by averaging the RMS obtained at each LST bin.


\section{Validation and Refinement}
\label{sec_validation}
\begin{figure}[h!]
\centering
\includegraphics[width=\textwidth]{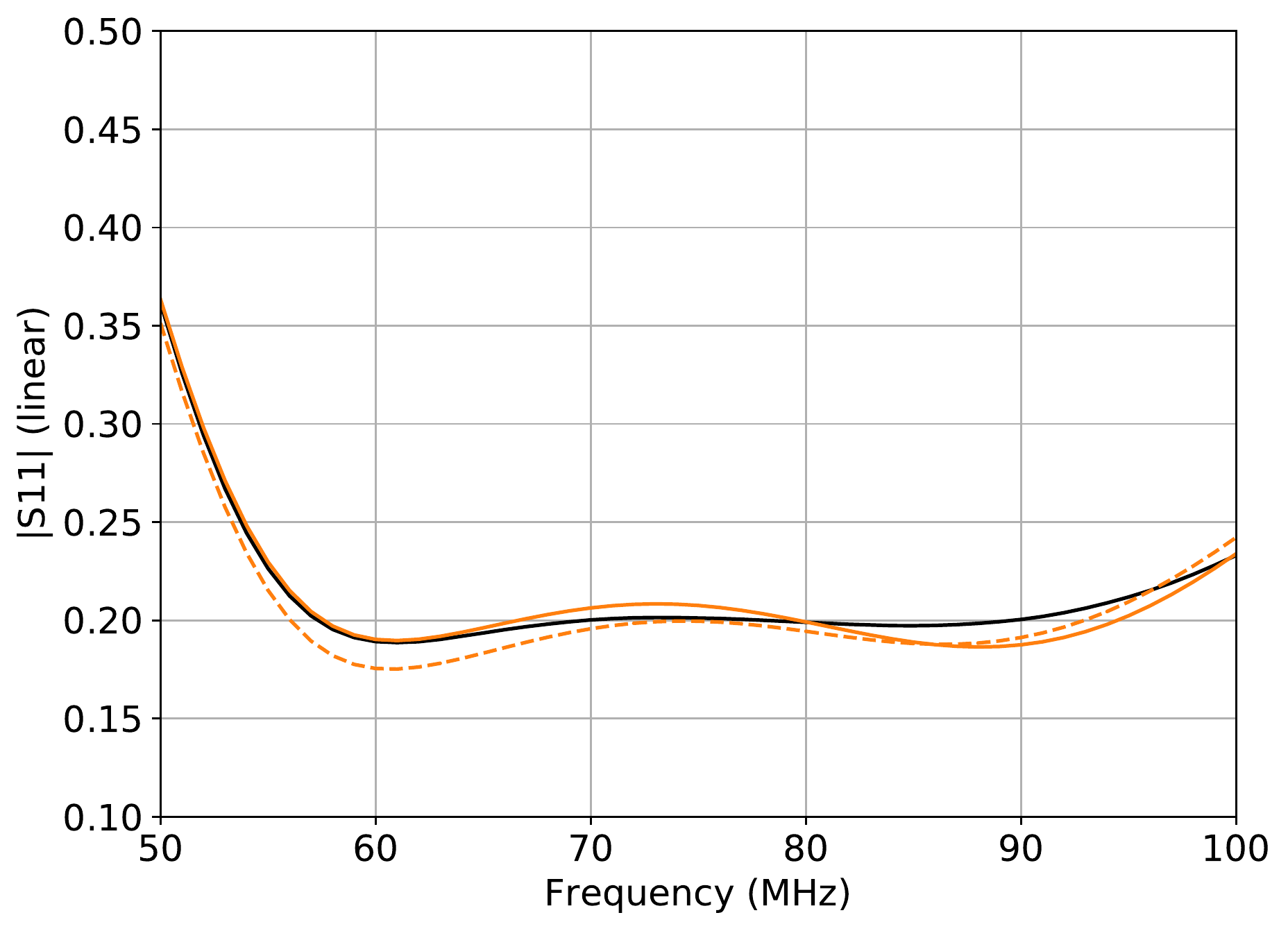}
\caption{Simulated antenna reflection coefficient (S11) for the low-band antenna over the 10~m~$\times$~10~m ground plane and MRO soil below from FEKO (orange solid curve) and from HFSS-IE (orange dashed curve) compared to one day of calibrated data (2015-12-08) measured in the field (black solid curve). The agreement between the simulations and measurement is within 0.015.}
\label{S11}
\end{figure}

While we are not able to measure the beam pattern of the deployed antenna, we can measure its S11 as a function of frequency.  Thus, to gain confidence that the simulations capture the actual properties of the low-band antenna, we compare its measured S11 with the simulated result from two different solvers, FEKO and HFSS-IE. All three software packages generate the same S11 values, within 0.015 (linear units), for our antenna model. This comparison is shown in Figure~\ref{S11}, where we see that the agreement in the absolute value of the S11 between the simulations and the measurement is within about 0.01 (limited by the adaptive meshing accuracy of the solvers). For absolute calibration of EDGES measurements for 21~cm cosmology, the antenna reflection coefficient must be known with an accuracy of 0.0003 (equivalent to about 0.01~dB) or better \citep{Monsalveone}.  We see from Figure~\ref{S11} that simulations do not provide sufficient agreement with the deployed antenna to rely solely on the modeling. We find that including the real ground and ground plane in the model still does not provide sufficient agreement to eliminate the need for in-situ antenna S11 measurements.  Nevertheless, the simulations show broad agreement with the deployed antenna indicating they do capture the primary properties of the antenna.

\subsection{Ideal Infinite PEC Ground Plane}
\label{PEC_ground}
We begin by cross-checking the results from the three packages using a simplified antenna model.  Modeling the actual ground plane realistically is challenging so we first consider a model with an idealized ground plane.  The antenna is modelled with its exact dimensions presented in Table~\ref{table-antenna-characteristics}, but instead of using the actual ground plane dimensions and soil properties, we model the ground plane as an infinite PEC.

\begin{figure}[h!]
\centering
\includegraphics[width=\textwidth]{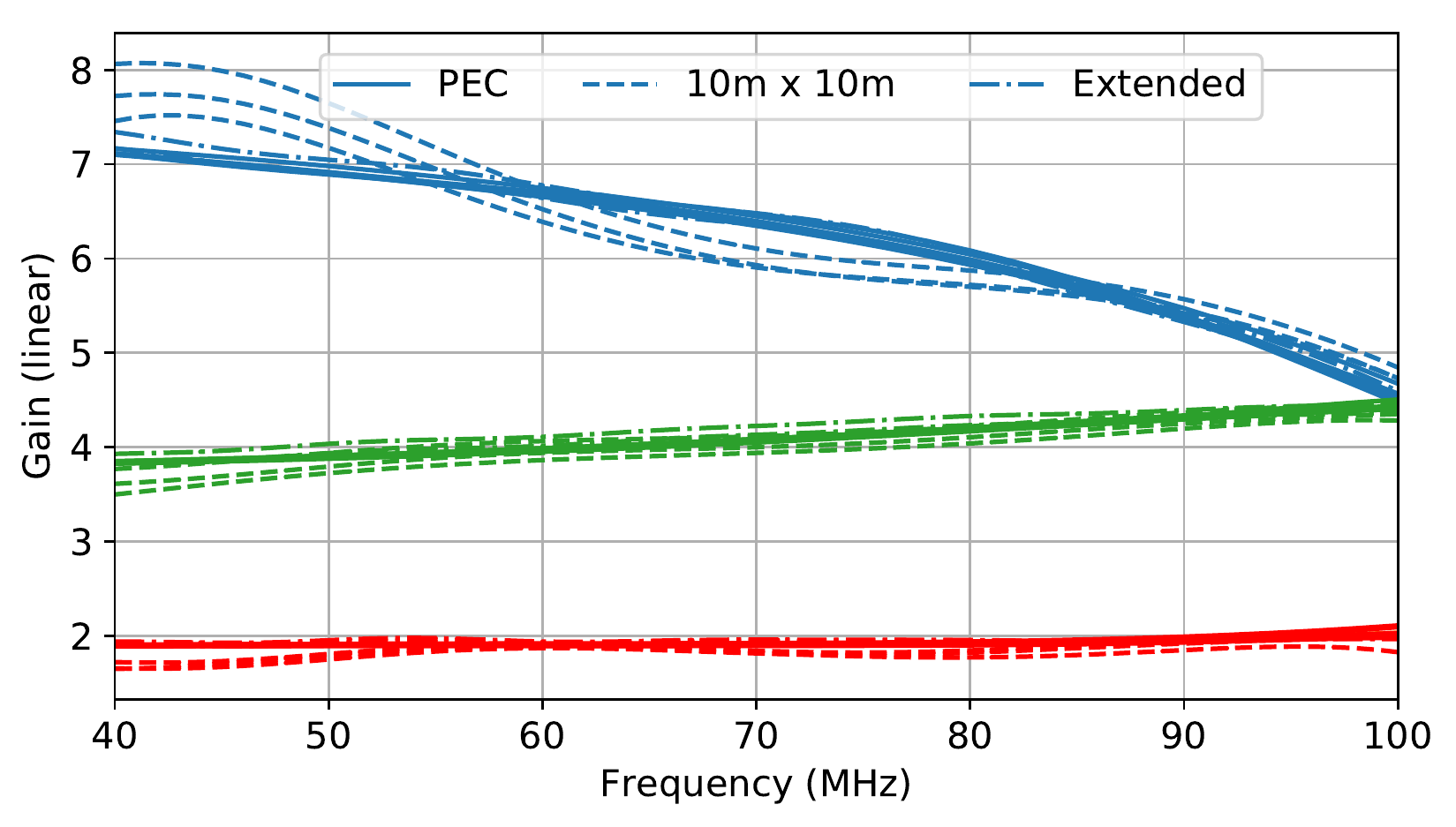}  
\caption{Antenna beam gain as a function of frequency for the three ground plane cases: PEC (solid), 10~m~$\times$~10~m (dash), and extended (dash-dot). Each ground plane case is shown at three viewing angles: zenith (blue), 45$^\circ$ elevation along the E-plane (red curves), and 45$^\circ$ elevation along the H-plane (green curves). The gain curves from all of the EM solvers are shown (but not labeled individually). In most cases the solutions from the different EM solvers overlap at the resolution of this plot, although some differences are apparent in the 10~m~$\times$~10~m solutions. As expected, the PEC and extended ground plane cases are more similar. } 
\label{all-PEC-low-punc}
\end{figure}

The far-field gain patterns obtained from each of the solvers are analysed using the metrics discussed in Section~\ref{sec_fom}. The antenna gain variations along frequency for a few viewing angles on the beam are plotted in Figure~\ref{all-PEC-low-punc} for each of the different solvers. As we can see, the values are found to be similar for each of the solutions. In all cases, the gain at 45$^\circ$~elevation is seen to be lower in the excitation plane (E-plane) compared to the perpendicular plane (H-plane). This is as expected because the total physical length of the antenna is longer along the excitation axis, implying the beam width will be narrow along the E-plane. 

\begin{figure*}
\centering
\includegraphics[width=\textwidth]{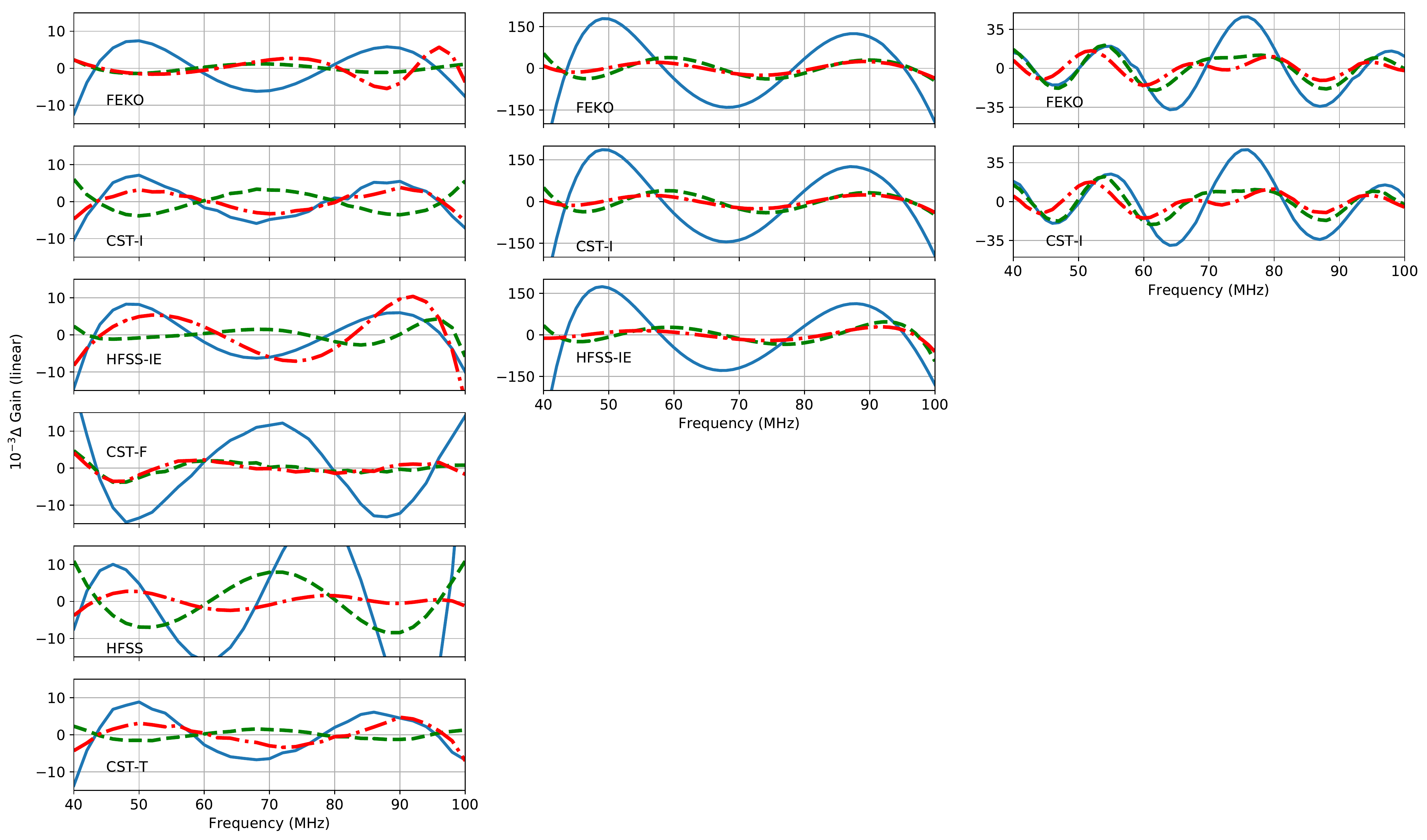}
\caption{Residuals after fitting and subtracting a regular three-term polynomial in frequency to the antenna beam gain along three example viewing angles for each solver and ground plane case: [left] PEC, [middle] 10~m~$\times$~10~m and [right] extended ground plane. For each model all the available simulation solutions are shown. The residuals are calculated for the gain curve at zenith (blue solid), 45$^\circ$ elevation along the E-plane (green dash), and 45$^\circ$ elevation along the H-plane (red dash-dot). The maximum level of chromaticity for the PEC ground is the lowest among all the three cases at a level of $\approx$ 0.010. The 10~m~$\times$~10~m shows a maximum deviation of 0.15 and the extended shows an improvement with the level of the residuals of $\approx$ 0.040.}
\label{all-PEC-low-res}
\end{figure*}

\begin{figure}[h!]
\centering
\includegraphics[width=\textwidth]{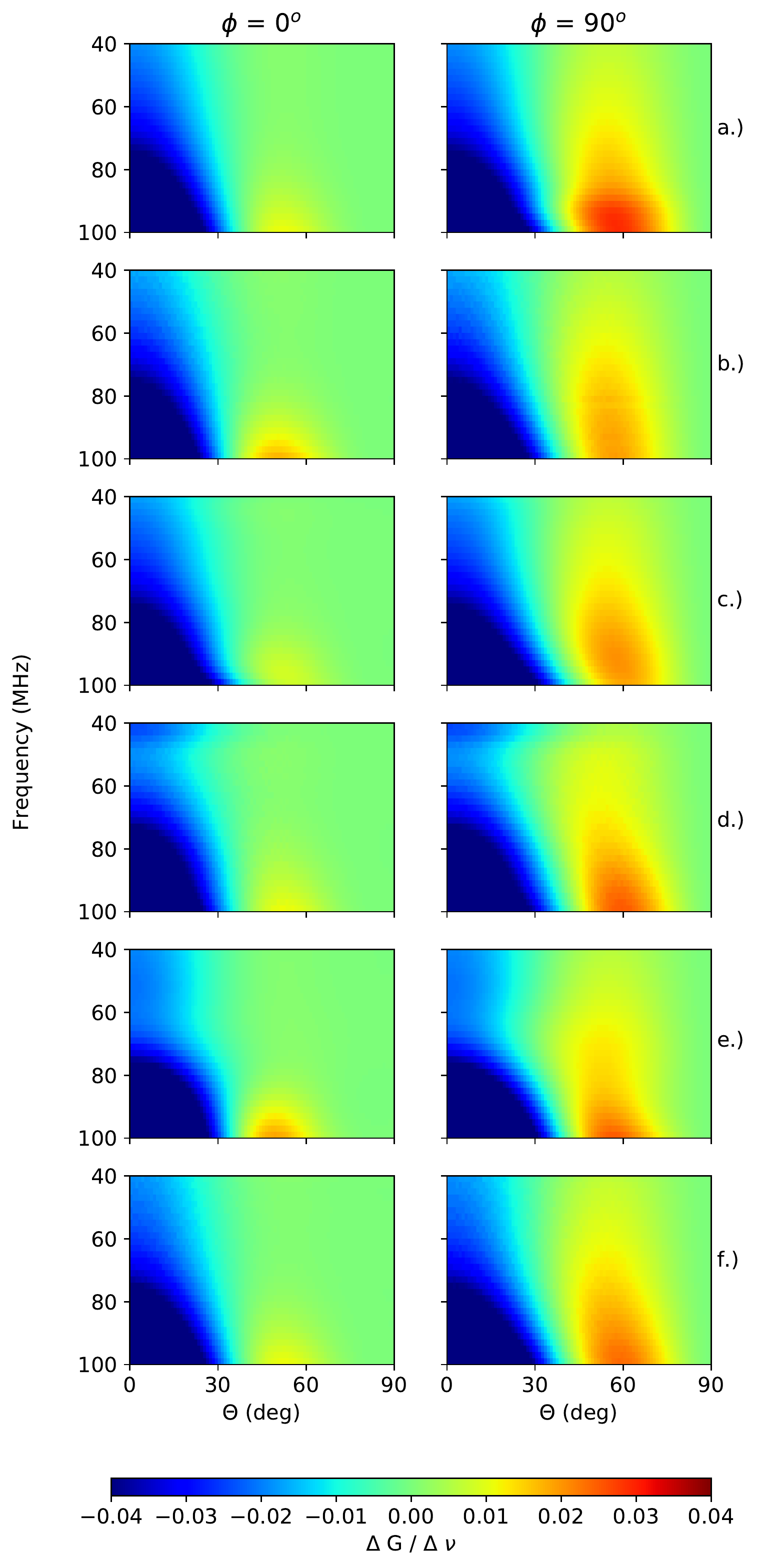}
\caption{Change in gain per MHz as a function of elevation angle along the E-plane (left) and H-plane (right) for the EDGES low-band dipole placed over an infinite PEC ground.  Each row shows the result from a different solver: a.) FEKO, b.) CST-I, c.) Filtered HFSS-IE solution, d.) CST-F, e.) HFSS and f.) CST-T.}
\label{all-PEC-low}
\end{figure}

The left panel of Figure~\ref{all-PEC-low-res} shows the residuals that characterize the beam chromaticity of the EDGES low-band antenna over a PEC ground for all solvers. We see that the residuals are similar between solvers, in particular those that implement the MoM technique. The maximum deviation of the gain from a regular three-term polynomial fit is found to be about 0.008 (linear units) for all cases except from the FEM solvers, in which case it is about 0.015 (linear units). The plots in Figure~\ref{all-PEC-low} show the frequency derivative of the gain at specific viewing angles as introduced in Equation~\ref{eqn_derivative}. We see that the gain derivative plots from  FEKO, CST-I and HFSS-IE are qualitatively similar. 



\subsubsection{Compensating for Fine Structure in HFSS-IE Models}
\label{sec_HFSS-IE}
\begin{figure}[h!]
\centering
\includegraphics[width=\textwidth]{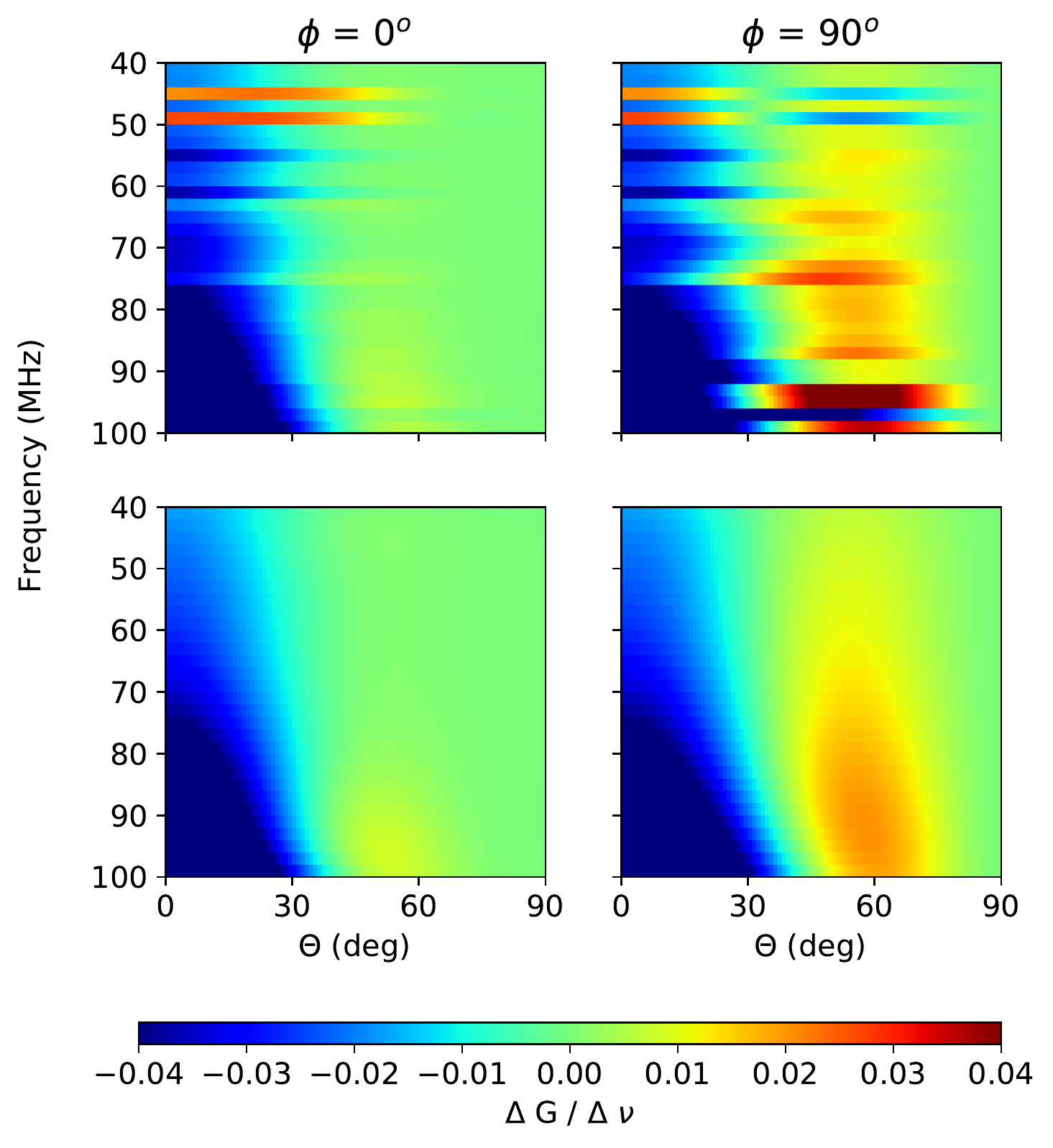}
\caption{Change in gain per MHz as a function of elevation angle along the E- and H-plane for the blade dipole placed over an infinite ground (PEC) modeled using: [top] HFSS-IE solver and [bottom] HFSS-IE solution with a regular 6-term polynomial fit.}
\label{MOM-PEC-low-HFSS}
\end{figure}

We found high-order variations with frequency in the residuals from the HFSS-IE solver (see Figure~\ref{MOM-PEC-low-HFSS}) and attribute them to numerical artifacts due to a lower accuracy of the solver. To compensate for this excess structure, we fit a regular 6-term polynomial ($\sum_{n=0}^{n=6} p_n x^n$) in frequency at every sampled azimuth-elevation point in these solutions and use the resulting fits in place of the original solutions. The HFSS-IE panels in Figures~\ref{all-PEC-low-punc}, \ref{all-PEC-low-res}, and~\ref{all-PEC-low} show the results after following this procedure. They qualitatively match the solutions from the other solvers. Other than the raw output from HFSS-IE shown in Figure~\ref{MOM-PEC-low-HFSS}, the analyses of all HFSS-IE beam patterns presented in this manuscript have been performed on polynomial-filtered beam patterns.

\subsubsection{Radiation Box and Meshing Refinement in FEM Models}

\begin{figure}[h!]
\centering
\includegraphics[width=\textwidth]{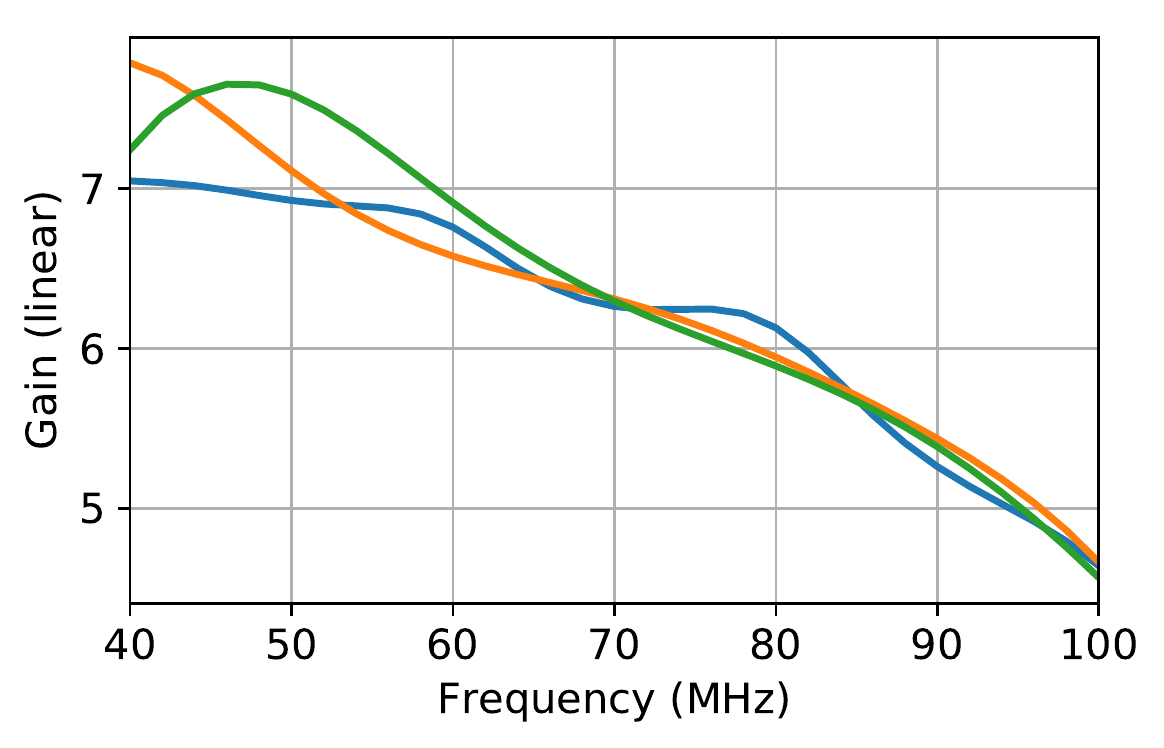}
\includegraphics[width=\textwidth]{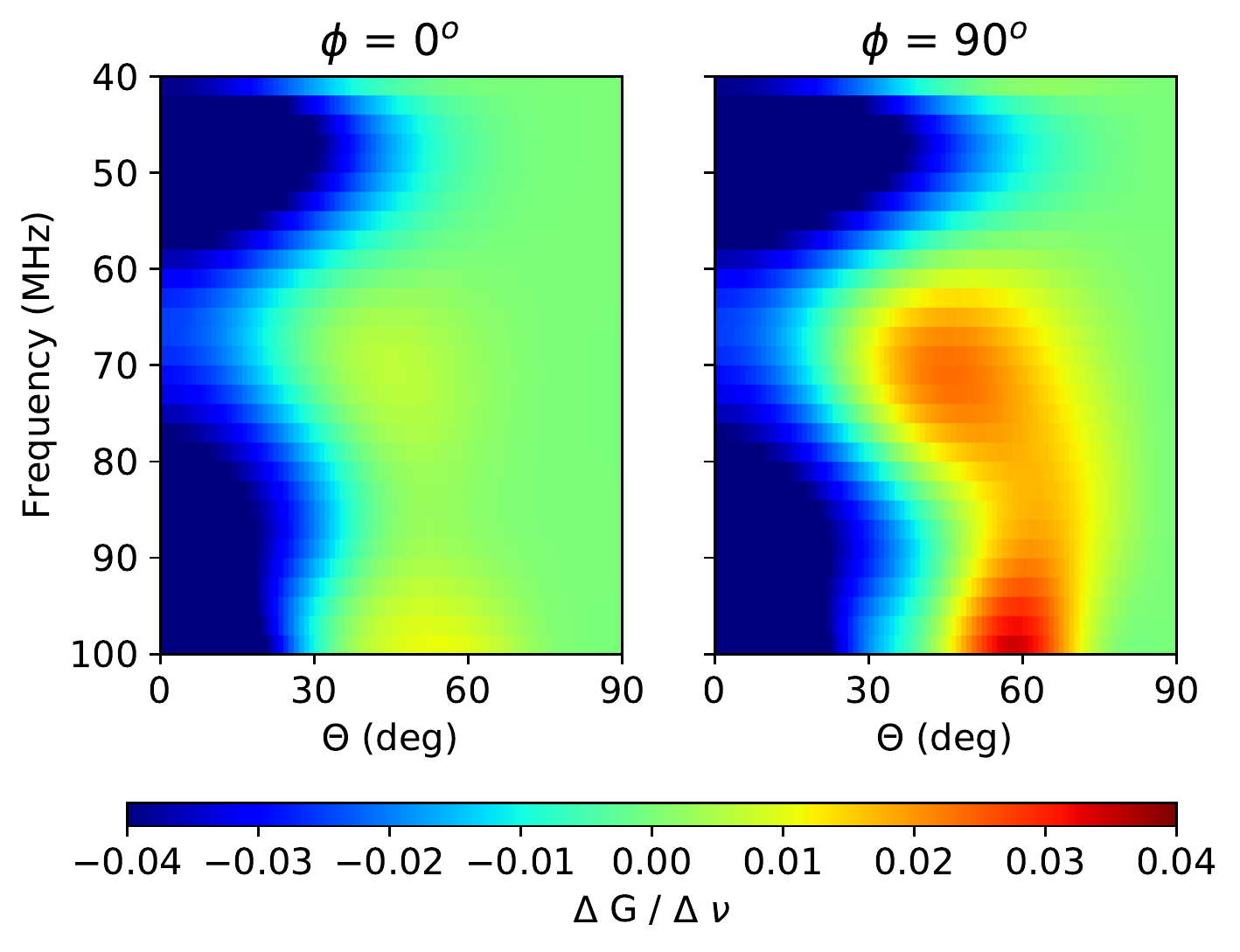}

\caption{[Top] Zenith gain as a function frequency obtained from HFSS simulations for different radiation box sizes with Absorbing Boundary Condition. The blue, orange and green curves represents $\frac{\lambda}{1.5}$, $\frac{\lambda}{3}$ and $\frac{\lambda}{4}$ box paddings, respectively. The gain varies with box size and doesn't converge even for a large value of padding such as $\frac{\lambda}{1.5}$. [Bottom] Change in gain per MHz as a function of elevation angle along the E-plane (left) and H-plane (right) for the blade dipole placed over an infinite PEC ground below and a surrounding ABC type radiation box which is modeled using HFSS solver. This gain variation is larger than any of the solutions shown in Figure \ref{all-PEC-low}.}
\label{HFSS-FEM-PEC-low}
\end{figure}

In HFSS, the antenna was initially simulated using the Absorbing Boundary Condition (ABC) on the radiation box with different padding distances from each of the model edges. On varying the radiation box size, the gain was also seen to vary (Figure~\ref{HFSS-FEM-PEC-low}). 
This dependence of the gain solution on the radiation box size, combined with larger chromaticity than any of the MoM solutions, suggested that the effects were caused by the solver and not inherent in the antenna.  This issue was solved by using a Perfectly Matched Layer (PML) boundary setting instead of the ABC radiation box. A PML boundary ensures nearly zero reflection at the radiation boundary, making the far-field solutions independent of the box size. The solutions obtained after using a PML boundary are used in Figures~\ref{all-PEC-low-punc}, \ref{all-PEC-low-res}, and~\ref{all-PEC-low}. This solution looks qualitatively similar to the MoM solutions.  The CST-F solver also requires a bounding box around the EM structure. With the same reasoning as noted for the HFSS simulation, a PML boundary was used in the CST-F simulation. 

The solutions from the CST-F models were further seen to vary based on the initial mesh setting. This was the case even though we adopted adaptive mesh settings, as we did with every solver. We increased the mesh cells until the obtained solutions were independent of the mesh size. In CST-F, this was achieved at a higher mesh density of $\lambda/20$ seed mesh size compared to $\lambda/8$ for the HFSS model. The factor of 2.5 difference which implies 2.5$^3$ more cells in CST-F and that the latter required a bounding box padding of $\lambda/3$ (vs $\lambda/4$ used by HFSS) resulted in the highest number of mesh cells ($\sim$ 4.6 Million) used for the CST-F simulation among all solvers. This also translated to needing 16x more memory. The  analyses  of CST-F  beam  patterns presented  in  this  manuscript  have  been  performed on simulations with the PML boundary and higher seed mesh setting. In Table~\ref{tabletwo} we summarize the performance properties of each of the EM solvers as configured for our models. The corresponding gain derivative plot is shown in Figure~\ref{all-PEC-low} and is in agreement with the solutions from the other solvers.

We also note that the FDTD solver, CST-T, uses a PML boundary for the radiation box by default for determining the far-fields. Like the CST-F solver, this solution method also required a denser seed mesh setting. This was expected as the FDTD technique is known to be computationally intensive in terms of memory and time \citep{1011769}.





\subsection{Best-Case FoM}
\label{best-case_fom} 
\begin{figure}[h]
\centering
\includegraphics[width=\textwidth]{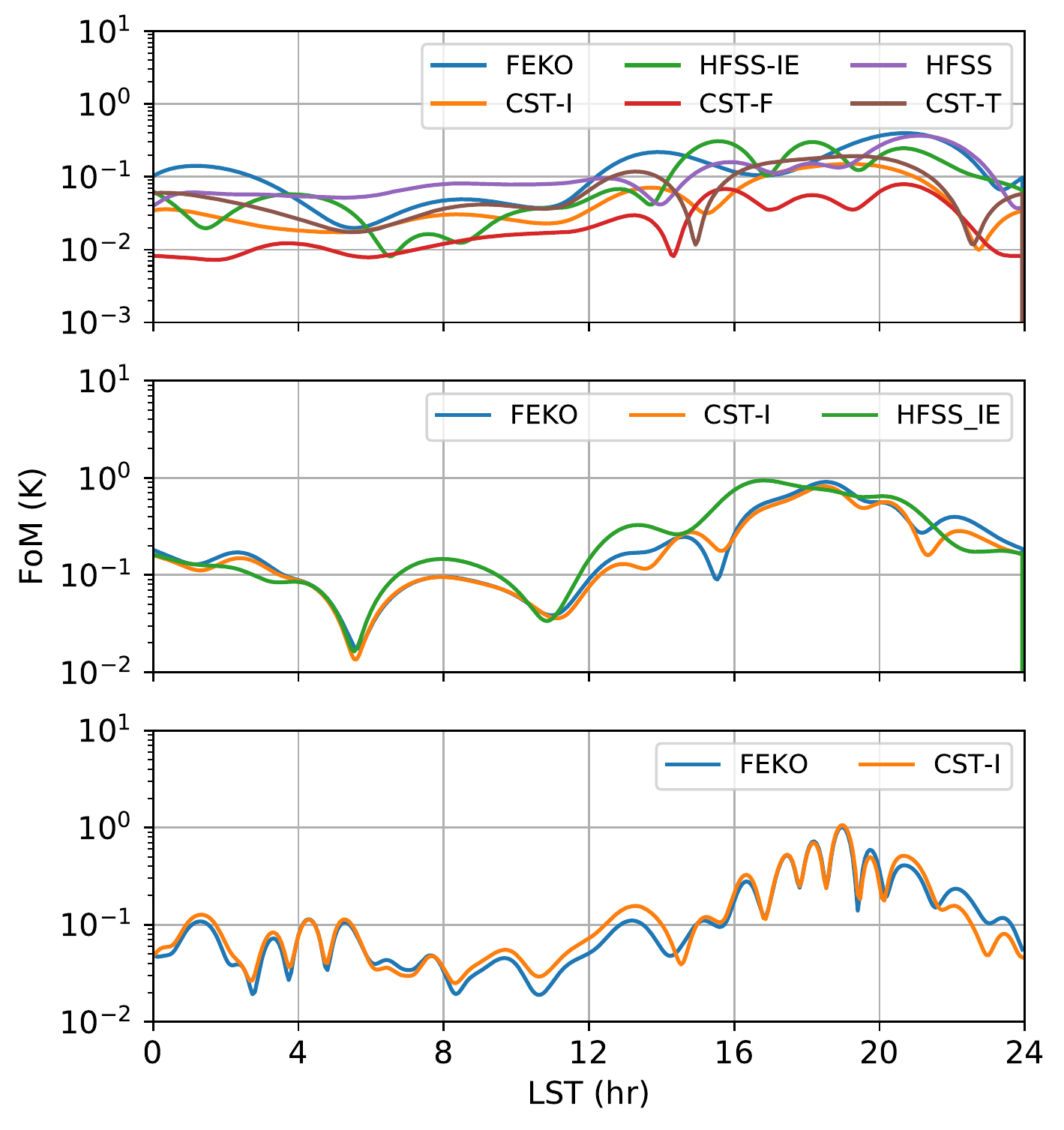}
\caption{FoM as a function of LST for the three ground plane cases: PEC (top), 10~m~$\times$~10~m (middle) and extended (bottom).  All the available beam solutions from different solvers are shown. The FoM was calculated with Equation \ref{eqn-fom} and using a five-term LinLog model for the spectrum (Equation \ref{linpoly}).}
\label{fom-new}
 \end{figure}

For the idealized infinite PEC ground plane test model, the FoMs derived from each of the EM solvers are shown in Figure~\ref{fom-new}. They all follow a similar trend, each peaking when the Galactic center is near zenith (LST$\approx$18~hr) and reaching a minimum when the Galactic center is not visible or is low on the horizon (LST$\approx$6~hr). For all the solutions, the FoM reaches a minimum of about 40~mK. The solvers all yield a maximum when the Galactic center is in the beam, although the values range from 100 to 500~mK and the relative position of the maximum varies. 

In Table~\ref{tabletwo}, we show the LST-averaged FoM for each solver. We see that solutions from all the solvers and methods except for those from HFSS, are in good general agreement and do not vary by more than 25$\%$. The beam solutions from HFSS indicate higher chromaticity. Given the similarities between the other solvers, this can be attributed to a low numerical accuracy of HFSS, and in particular of HFSS-IE (as noted in 3.1.1). We also see that the chromaticity level of 0.01 (shown in Fig. \ref{low-band-ex-residues}) of the EDGES low-band blade antenna over PEC ground results in an antenna temperature residual spectral structure of $65$~mK on average and as good as $<$30~mK at some LSTs (Figure \ref{fom-new}). This generally meets the needs for 21~cm cosmological experiments searching for a signal of order $\approx$ 100~mK, even without applying any beam corrections during analysis. While the idealized infinite PEC ground plane yields a best-case FoM given the EDGES antenna design, we will see in the next section that the extended ground plane performs nearly as well.


\begin{table*}[ht]
 \caption{Comparison of EM solvers}
 \centering
 \begin{tabular}{c c c c | c c c}
 \hline
  Solver & Technique &   Mesh cells\footnote{Determined for ideal ground plane (PEC)\label{f1}}  & Memory $^{\rm{a}}$ &   \multicolumn{3}{c}{LST-averaged FoM\footnote{For the real ground cases, the FoM after applying a beam correction to the simulated and actual observations is indicated in parentheses.  The beam correction is calculated as in \citet{Mozdzen_2016,Mozdzen_2019} using the sky model and the FEKO model for all cases.  As expected, the beam correction removes all of the residual structure when applied to the simulated observations from the FEKO beam model.  It also substantially lowers the FoM for the simulated observations generated by the other solvers, indicating their beam patterns are similar to the FEKO patterns.  When applied to the actual data, we see that the beam correction lowers the FoM by 33\% for the 10~m x 10~m ground plane and 8\% for the extended ground plane. }}  \\
  
         &           &            &    & PEC\footnote{Section~\ref{PEC_ground}} & {10~m~$\times$~10~m\footnote{Section~\ref{org_ground}}} & { Extended\footnote{Section~\ref{gp-imp}}} \\
         
          &           &           &    &  & {ground plane} & {ground plane}\\
         &           &   (\#)         &  (MB)  & (mK) & {(mK)} & {(mK)} \\
 \hline
   FEKO    & MoM  & 2,262   & 103       & 61.8  & 240 (0) & 82 (0)\\
   CST-I   & MoM  & 5,964  & 1,556     & 63.3  & 212 (54) & 101 (40)\\
   HFSS-IE & MoM  & 5,210   & 310     & 109.0  & 292 (77) &---\\
   CST-F   & FEM  & 4,654,080 & 16,055     & 50.7  & ---&--- \\
   HFSS    & FEM  & 15,102 & 1,231     & 84.0 & --- &---\\
   CST-T   & FDTD & 926,478   & 1,257     & 68.0  & ---&--- \\
   Actual Data & --- & --- & --- & --- & 280 (188) &  130 (120) \\
 \hline
 \label{tabletwo}
 \end{tabular}
 \end{table*}

\section{Results}

In Section~\ref{sec_validation}, we verified that the simulated beam pattern and chromaticity of the blade antenna over an ideal PEC ground plane is consistent between the three different EM simulation techniques and three independent software packages. However, our idealized model with an infinite PEC ground plane represents a best-case scenario.  In reality the ground plane will have a finite size and the currents will reach the actual soil below the ground plane, creating additional frequency-dependent structure in the beam patterns. We now proceed to simulate realistic soil below the antenna structure along with the actual, finite ground planes used by EDGES. The EM solvers that have the capability to simulate an infinite dielectric ground are those that are based on the MoM technique. They are FEKO, HFSS-IE and CST-I. 

\subsection{Realistic Soil and 10~m~$\times$~10~m Ground Plane}
\label{org_ground}
We begin by modeling the original ground plane for the low-band antenna, which was a 10~m~$\times$~10~m square made from metal wire mesh, as described in Section~\ref{sec_ant-model}.  To reduce the complexity of the EM simulations, we model the ground plane as a finite PEC solid plate, rather than as a mesh.  The soil below the ground plane is included as a half-space boundary condition (i.e, it is infinite in $\pm X$, $\pm Y$, and $-Z$~directions).  In this case, the MoM solver engine uses a different Green's function kernel for the lossy dielectric media in the $-Z$ hemisphere than for the free-space above.  For the soil properties, we use a conductivity of $\sigma = 0.02$~Sm$^{-1}$ and a relative permittivity of $\epsilon_r = 3.5$. These values were measured by \citet{Sutinjo_2015} for samples of soil from the MRO.

\begin{figure}[ht!]
\centering
\includegraphics[width=\textwidth]{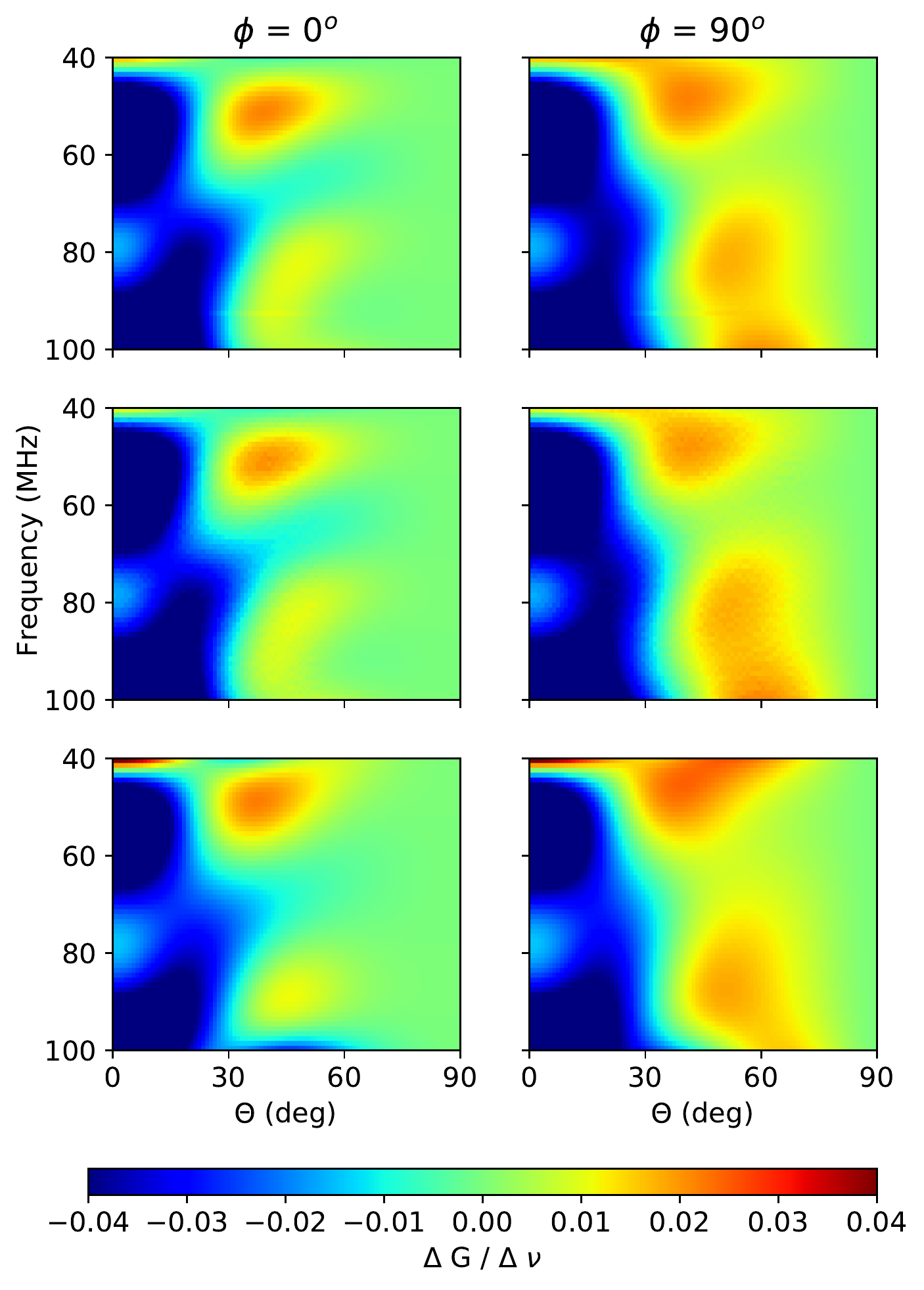}
\caption{Same as Figure \ref{all-PEC-low}, but for the 10~m~$\times$~10~m ground plane case and using only the Method of Moments solvers: FEKO, CST-I and, HFSS-IE. This gain variation plot is seen to have more structure compared to the PEC case.}
\label{MOM-dielectric-low}
\end{figure}


The solutions for the derivatives of the gain across frequency are plotted in Figure \ref{MOM-dielectric-low}. In the plots we see that the realistic finite ground plane solutions have more structure in frequency than in the ideal ground case (Figure \ref{fom-new}).  There is good qualitative similarity in the beam chromaticity between the three real ground simulations, thus giving confidence that this structure is real. 
The dashed lines in Figure~\ref{all-PEC-low-punc} shows the absolute gain variations along a few viewing angles on the beam patterns. The residuals to a three-term polynomial across frequency to each of the curves for the finite ground plane model show larger fluctuations than the ideal ground case (see Figure \ref{all-PEC-low-res}). The maximum peak-to-peak residual across frequency is about 0.2 (linear units), compared to 0.008 for the ideal ground. Good agreement is seen between the solutions from the three solvers.  Quantitatively, the higher chromaticity results in a larger value for the FoM compared to the ideal ground case, as can be seen in Figure~\ref{fom-new}.
The FoM peaks at about LST$\approx$18~hr with a value of 1~K and drops to 20~mK at LST$\approx 5.5$~hr. The LST-averages of these FoM values are summarized in Table~\ref{tabletwo} and are of the order of $\approx$ 250~mK which is about 3.5~times larger than the ideal ground case. CST-I and FEKO solutions agree to 13.4$\%$. Similar to the PEC case, the HFSS-IE result is significantly higher.

\subsection{Extended Ground Plane}
\label{gp-imp}
We showed above that the original square 10~m~$\times$~10~m ground plane used for the EDGES low-band instrument resulted in the LST-averaged FoM for the simulated antenna spectra being in the range 212-292~mK, whereas we expect the lowest chromaticty that we could achieve with a blade planar dipole should approach the $\approx$ 65~mK limit of the ideal ground case shown in Section~\ref{sec_validation}. Improving the performance of the ground plane beyond its original design was important to meet the science objectives.  With the objective of achieving an LST-averaged FoM below 100~mK, we conducted a design study aiming for a factor of two improvement. In order to decrease the chromaticity, we must reduce the contribution of the reflected waves from the soil to the main beam response. This involves reducing the currents that reach the soil, which can be done by increasing the ground plane size and/or modifying its shape. 

We simulated and analyzed different sizes and a few possible shapes of the ground plane using FEKO in order to identify designs that achieved our goal. In addition to simple square ground planes, we included plus-shaped designs and squares with triangular ``sawtooth'' extensions around the perimeter. Among the designs tested, we simulated plus-shaped ground planes of maximum length 15~m and 30~m. The 15~m and 30~m sized plus ground planes are made of five sections of 5~m~$\times$~5~m squares and 10~m~$\times$~10~m squares respectively. The sawtooth designs, with a central square and triangles along the perimeter, were similar to the extended ground plane described in Section \ref{sec_ant-model}, but with varying square sizes in the center. The plus-shaped and sawtooth configurations were considered based on the analysis done in \citep{Meng}, which showed that more edges along the perimeter of the ground plane reduces the reflection from the surrounding soil. The designs and shapes of the ground plane we explored were planar to avoid horizon response. The resulting FoM values are listed in Table~\ref{fom-trial}. 

Three main conclusions that can be drawn from the Table are: i) for a given shape, the larger the area, the lower the FoM, ii) for roughly the same area, the FoM of the sawtooth design is lower than the FoM of the plus shape, which in turn is better than the FoM obtained from the square ground plane and iii) for the same area of the sawtooth ground plane, increasing the number of edges beyond $16$ (by increasing the triangles used) did not affect the FoM significantly. A higher number of edges reduces the currents that reach the soil, which in turn helps to reduce the beam chromaticity. The plus-shaped and sawtooth ground planes also have an added advantage that they require less metal for a given tip-to-tip extent.

\begin{table}[h]
 \caption{Comparison of the Figure of Merit (FoM) for different cases of finite ground planes modeled with FEKO.  For the sawtooth designs, the size listed is the dimension of the inner square region.  The number of triangles added around the perimeter of the inner square is listed in parentheses.  The added triangles have 5~m base and 5~m height in all cases except for the bottom row, for which the triangles had 4~m base and 5~m height.}
 \centering
 \begin{tabular}{c c c c}
 \hline
  Size  & Shape & Area & LST-averaged   \\
  (m) & &  (m$^2$)&FoM (mK)\\
 \hline
   10 & Square & 100 & 213   \\
   20 & Square & 400 & 205  \\
   30 & Square & 900 & 192  \\
   15 & Plus & 125 & 255\\
   30 &  Plus & 500 & 181 \\
   10 (8) & Sawtooth & 200 & 151\\
   15 (12) & Sawtooth & 375 & 149\\
   20 (16) & Sawtooth & 600 & 80\\
   20 (20) & Sawtooth & 600 & 90\\
 \hline
 \label{fom-trial}
 \end{tabular}
 \end{table}
 
 \begin{figure}[h]
\centering

\includegraphics[width=\textwidth]{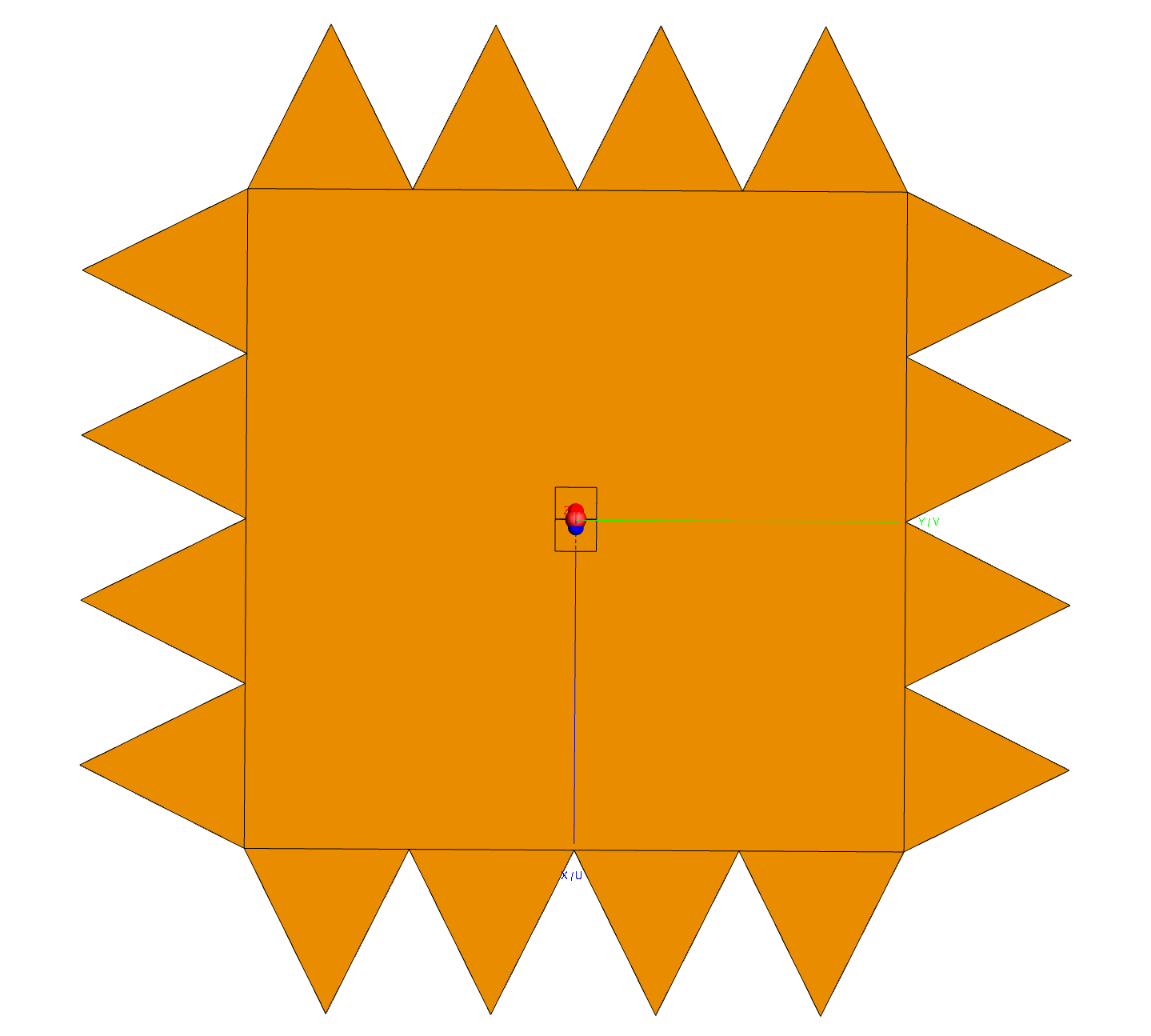}
\caption{Rendering of the extended ground plane.  It consists of a central square metal plane that is 20~m~$\times$~20~m.  Each side has four isosceles triangles that are each 5m at their base and 5 meters long to their tip.  The overall size of the ground plane is 30~m~$\times$~30~m. In the field, the width of each of the isosceles triangles is slightly narrower than 5 meters (4.8~m). This is due to the size of the mesh sheets we used to build the actual ground plane.}
\label{new-gnd}
\end{figure}

\begin{figure}[h!]
\centering
\includegraphics[width=\textwidth]{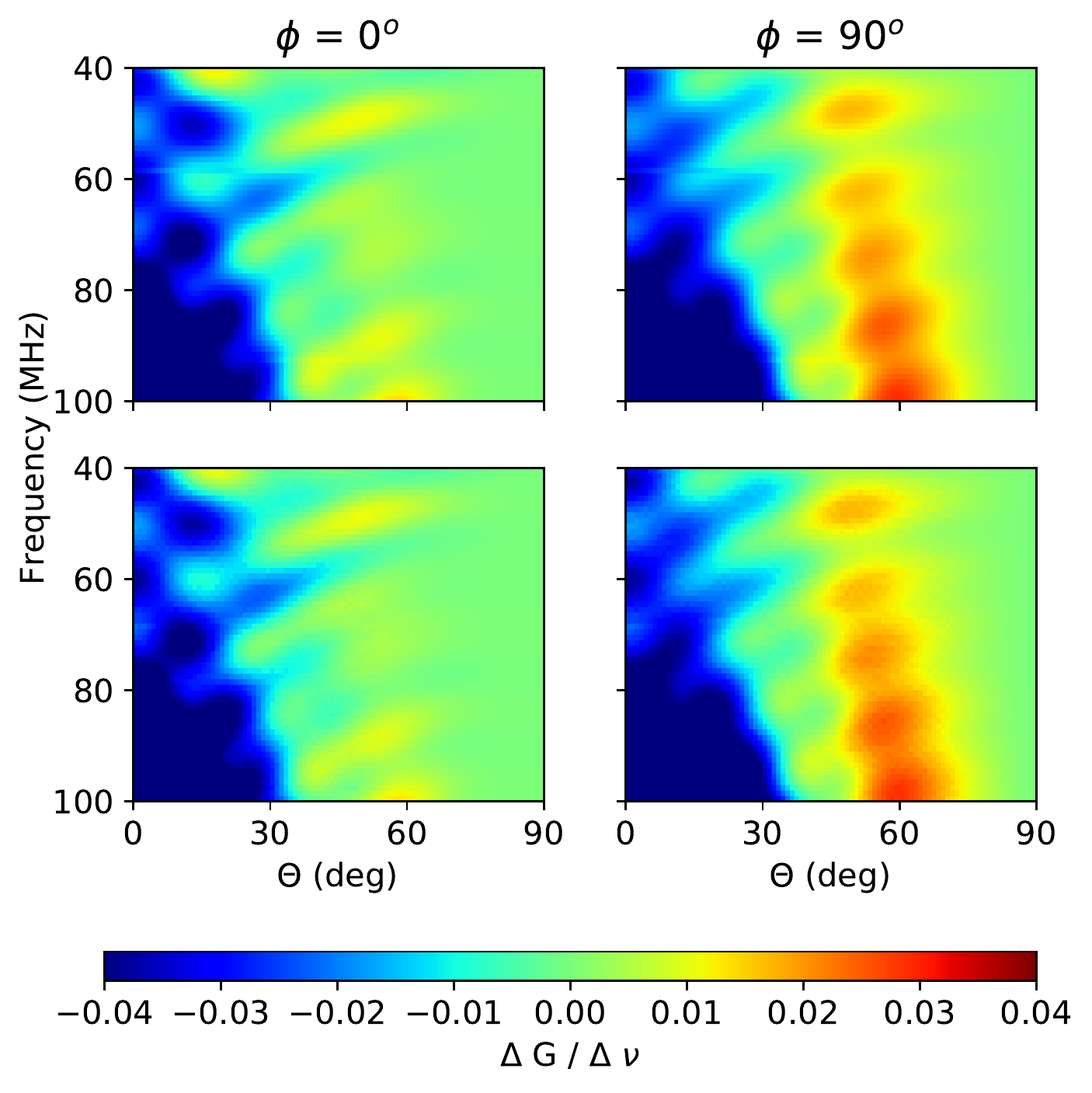}
\caption{Same as Figure \ref{all-PEC-low}, but for the extended ground plane case using the solutions from FEKO and CST-I. Along each viewing angle, the variation of the gain with frequency is more gradual compared to the 10~m~$\times$~10~m ground plane but larger than the PEC case.}
\label{MOM-dielectric-low-new}
\end{figure}

Based on this analysis, we arrived at the final design for the larger ground plane of 20~m x 20~m square with triangle extensions described in Section~\ref{sec_ant-model} and shown in Figure~\ref{new-gnd}. The extended ground plane was simulated in FEKO and CST-I.  We do not present solutions from HFSS-IE because the beam solutions could not be filtered with a 6-term polynomial in frequency to remove the solver errors (as described in Section~\ref{sec_HFSS-IE}) without smoothing structures that were part of the actual beam response. The beam derivative plots of FEKO and CST-I solutions are shown in Figure~\ref{MOM-dielectric-low-new}. We can see that, as expected, the extended ground plane produces more structure than the ideal ground, but at smaller frequency scales than the original $10$-m $\times$ $10$-m ground plane. Overall, the reduction of large-scale fluctuations in frequency improves the chromaticty. This is also shown in the right column of Figure~\ref{all-PEC-low-res}, where the residuals of the gain to a regular three-term polynomial are below 0.04 -- more than a factor of five improvement over the residuals produced by the original 10~m~$\times$~10~m ground plane. The solutions for the derivatives of the gain across frequency are plotted in Figure \ref{MOM-dielectric-low-new}. In the plots we see that the extended ground plane solutions have more structure in frequency than in the ideal ground case (Figure \ref{fom-new}) but more gradual compared to the 10~m~$\times$~10~m case. There is good qualitative similarity in the beam chromaticity between the FEKO and CST-I simulations, thus giving confidence that this structure is real. The FoM versus LST plots show that the RMS of the residuals never exceed 1~K for both FEKO and CST-I solutions, and is as low as 20~mK when the Galactic plane is below the horizon. 
The average FoM across LST with the extended ground plane obtained from FEKO is 82~mK and from CST-I is 101~mK (a difference of 22$\%$). The 10~m~$\times$~10~m and extended cases have roughly the same FoM when the Galactic center is near zenith, but the extended cases goes to a lower FoM at low LST for a longer period of time.

\subsection{Comparison with Data}
\label{data}
In the previous Sections, we have found that the beam solutions for four out of the six solvers tested agree to within 25\% for an ideal ground plane case and that the MoM solvers agree to within 20\% for finite ground plane cases.  Here we assess the overall accuracy of the beam solutions by comparing simulated observations to actual data.

\begin{figure}[h!]
\centering
\includegraphics[width=\textwidth]{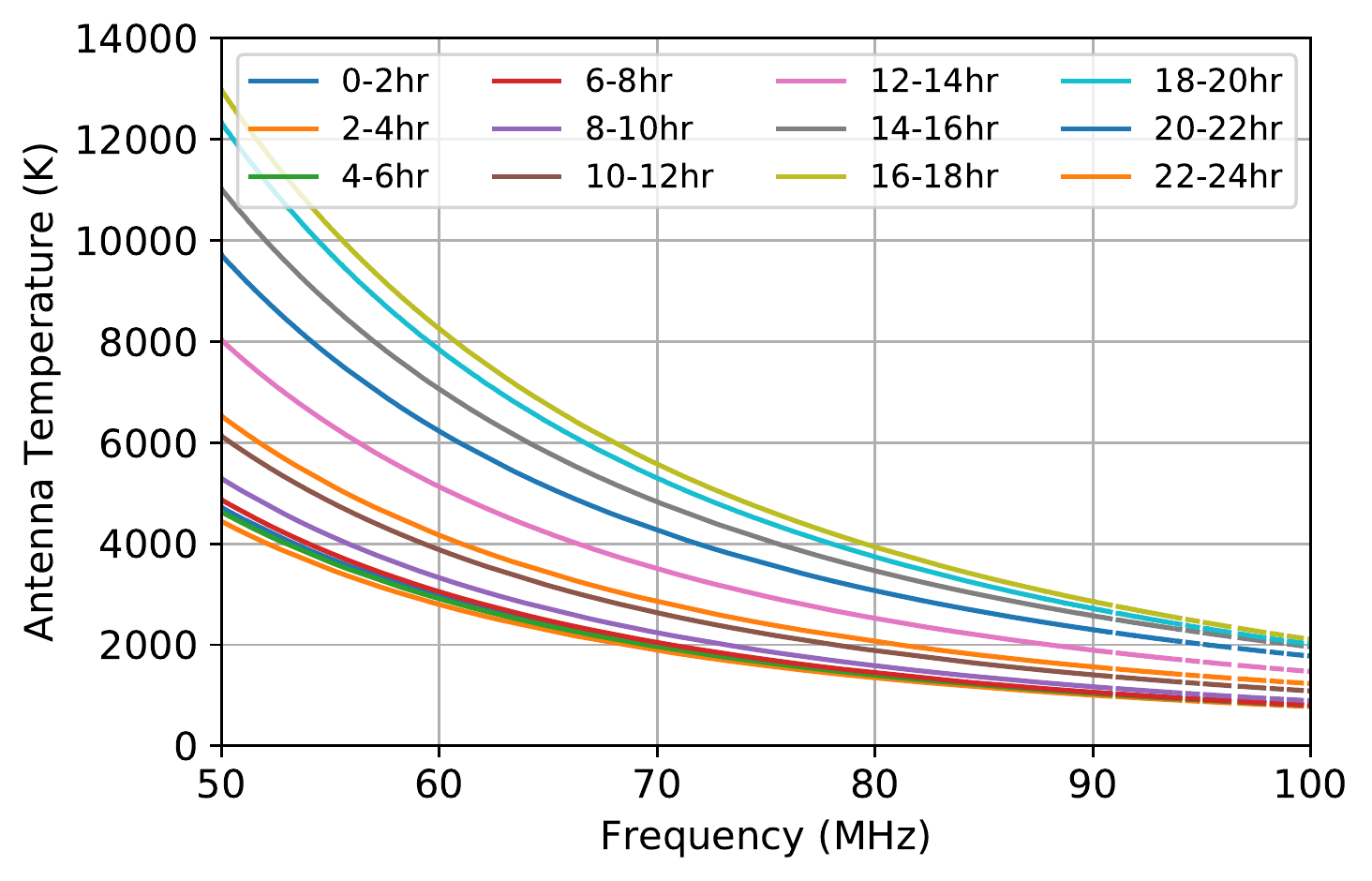}     
\caption{Observed antenna temperature as a function of frequency for the EDGES low-band instrument with a 10~m~$\times$~10~m ground plane.  Data are binned by LST into two-hour blocks. Some spectral channels between 90 and 100 MHz are omitted to avoid radio frequency interference.}
\label{data-bin}
\end{figure}

We use data collected from the EDGES low-band (low-1) instrument with both its original 10~m~$\times$~10~m ground plane configuration and its later extended 30~m ground plane. 
The data with the original ground plane spans the days 2015-286 to 2016-186 and with the extended ground plane 2016-258 to 2017-95. The raw data are cleaned for RFI and corrected for the calibrated receiver response. The data reduction pipeline we used here is similar to what was done with other EDGES data \citep{Monsalvetwo,bowman2018absorption}.
However, the main analysis here is carried out by comparing the observed spectra without any attempt to compensate for the chromatic beam effects to the simulated spectra obtained using the beam models. In order to provide the cleanest comparison possible, we omit daytime data and include only times when the Sun is at least 10 degrees below the horizon in order to avoid solar emission and reduce ionospheric interference. The processed data from each day are binned into two-hour blocks in LST and the data from all days were averaged together within their respective bins. The resulting binned spectra are shown in Figure~\ref{data-bin}. 

\begin{figure}[h!]
\centering

\includegraphics[width=0.81\textwidth]{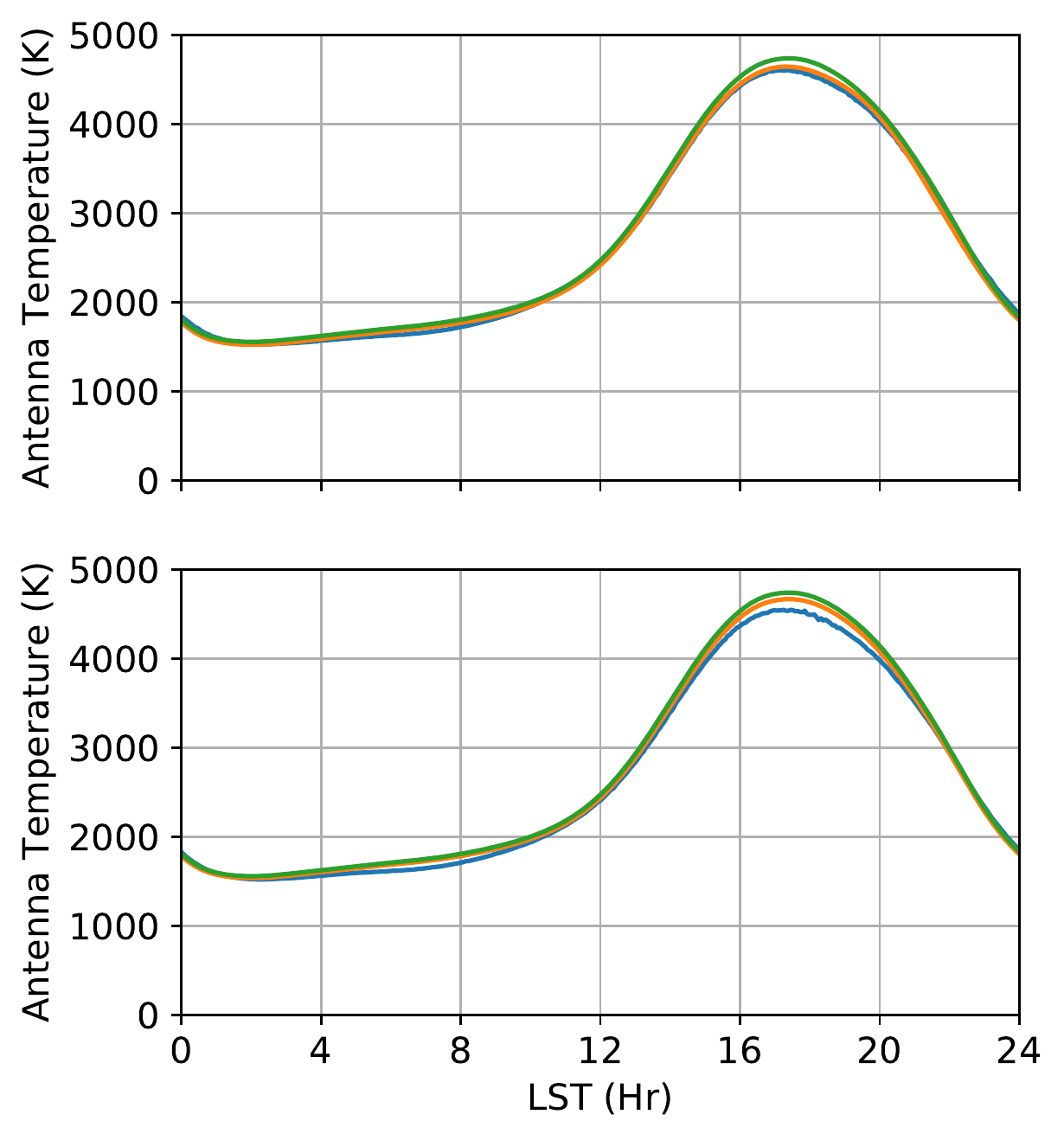}
\caption{Antenna temperature at 75 MHz as a function of LST for the observed data (blue), simulated spectra using FEKO beam patterns (green), and simulated spectra using CST-I beam patterns (orange). The top panel is for the 10~m~$\times$~10~m ground plane case and the bottom panel is for the extended ground plane.}
\label{driftscan-oldgp}
\end{figure}

\begin{figure}[h!]
\centering
\includegraphics[width=0.93\textwidth]{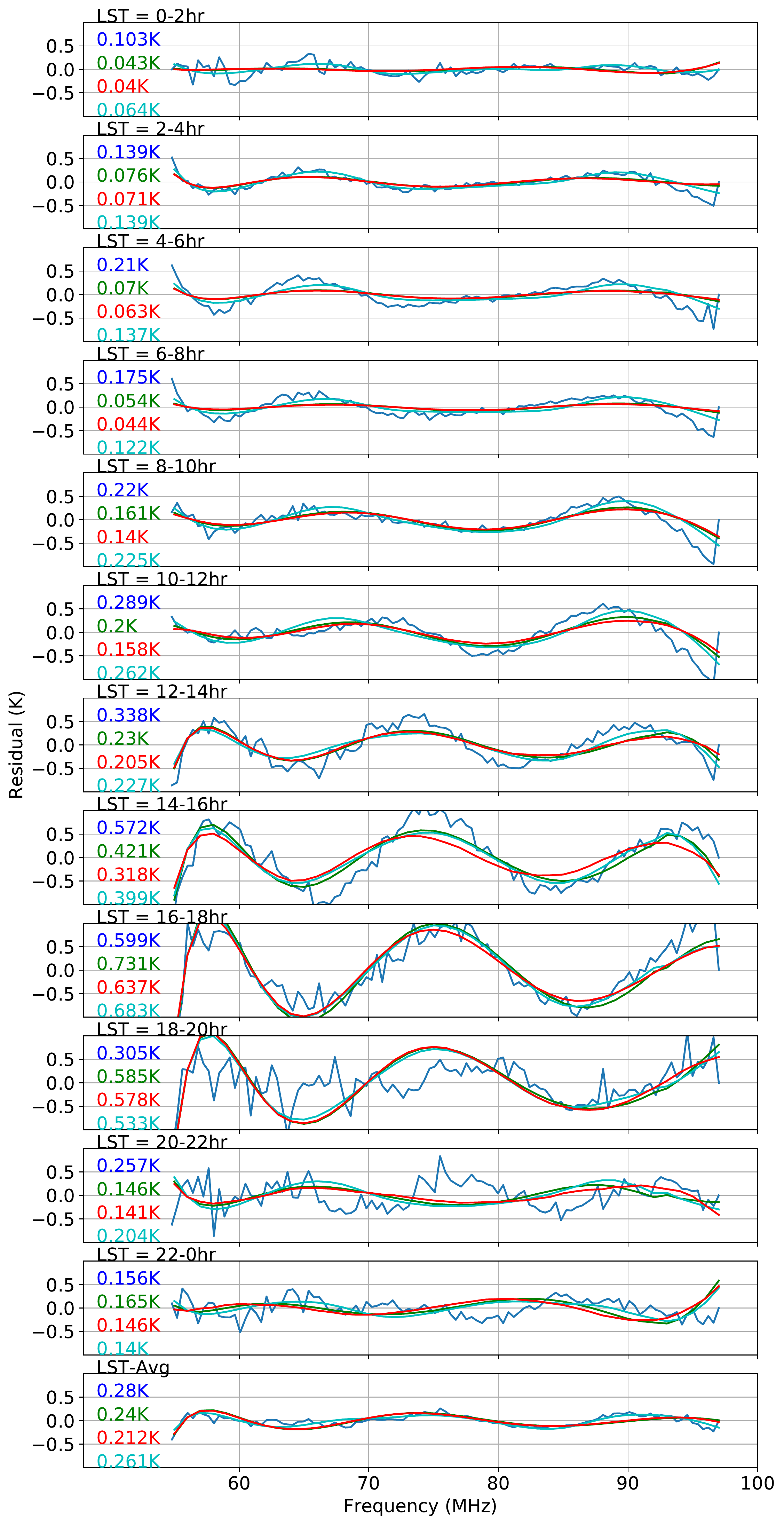}   
\caption{Residuals from the 10~m~$\times$~10~m ground plane after fitting and removing a five-term LinLog model to the data and the simulated data for each LST bin.  Each row corresponds to a different LST bin and shows the residuals to the data (blue), simulated data using the FEKO beam model (green), simulated data using the CST-I beam model (red), and simulated data including the absorption feature using the FEKO beam model (cyan) . The values displayed for each row indicate the FoM derived from the RMS of the residuals between 55 and 97~MHz.  The bottom row shows the residuals to the full 24-hour averages of the data and simulated data. The values listed for the bottom row are the LST-average FoMs, which differ slightly from the RMS of the average data because they are the average of the RMS values from each of the LST bins.}
\label{low-band-residues}
\end{figure}

For comparison to the data, we created simulated spectra following the method described in Section~\ref{sec_fom} using both the FEKO and the CST-I beam solutions. We averaged them into the same two-hour LST blocks as the data. To quickly validate the accuracy of the simulated spectra, we compare their predicted antenna temperatures at 75~MHz to the data in Figure~\ref{driftscan-oldgp}. The disagreement between the simulated spectra and the data varies with LST and is seen to be within 2$\%$ and 4$\%$ for the 10~m $\times$10~m and extended ground planes respectively. 
To see how well the spectra simulated using the beam models capture the low-level spectral structure in the observations, we fit the same five-term Linlog foreground model described in Equation~5 to the data and to the simulated spectra. 
We carry out the fitting over the frequency range of 55-97 MHz and then compare the residuals. This frequency band covers all the analyses cases carried out in \citep{bowman2018absorption}. 
In Figure~\ref{low-band-residues}, we show the 10~m~$\times$~10~m ground plane case. Looking across all LST bins, the residuals to the simulated spectra generally follow the trends seen in the residuals to the observations. They are high when the Galactic center is in the  beam and gradually decrease as the Galactic center moves to the horizon and below. The residuals in both cases reach a minimum ($\sim$100~mK) when the Galactic center is below the horizon.  Within each LST bin, we see that the residuals to the two simulated sky spectra capture the spectral variation of the the observation residuals well in terms of the phase of the peaks and troughs. We see the most significant disagreement between the data residuals and simulated residuals in the LST bin of 20-22~hr. This is attributed to the fact that the bright Galactic Center is in the portion of the beam where the gain is changing most rapidly. An error in the beam model is likely to have its strongest effect in the places where the gain changes most rapidly, both in frequency and in angle. This can be seen in Figure~\ref{MOM-dielectric-low} between 35 and 60 degrees elevation the derivative of the gain of the antenna varies the most with frequency. The differences between the simulated spectra and data are largest for the LSTs on either side of the Galactic Center transiting, but at 16-18~hr, the agreement is good again since the Galactic center is in the slowly changing center of the beam. The bottom plot in Figure \ref{low-band-residues} is the LST-average of all the residuals over the 24 hours. The average residuals align well between the simulations and observations. The FoM in the final subplot is the LST-averaged FoM as used in Table \ref{tabletwo}. The fractional difference between the FoMs from the simulations and the data is $\approx$ 15$\%$ for the FEKO solutions and $\approx$ 24 $\%$ for the CST-I solutions.

\begin{figure}[h!]
\centering
\includegraphics[width=0.93\textwidth]{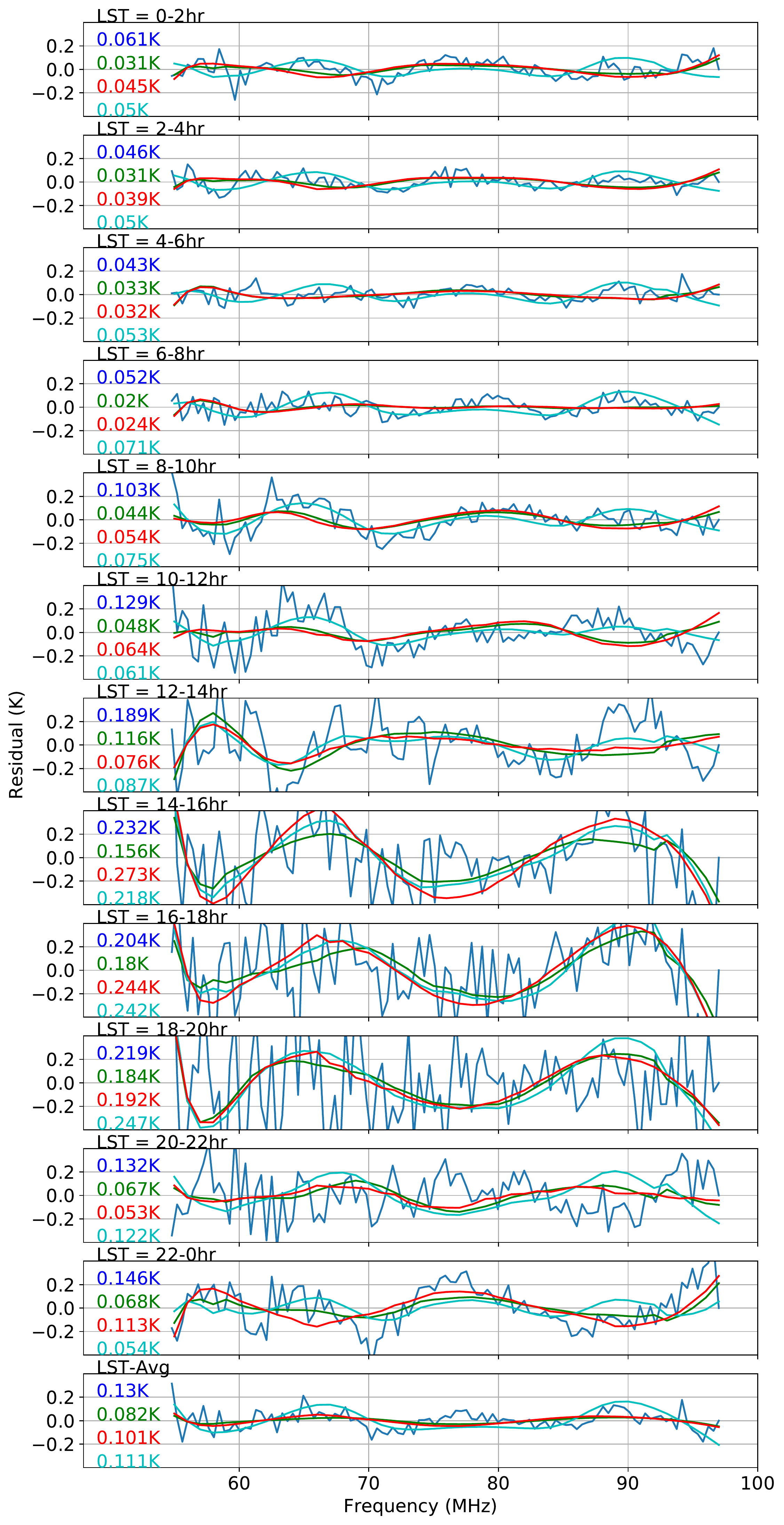}
\caption{Same as Figure \ref{low-band-residues}, but for the extended ground plane case}
\label{low-band-ex-residues}
\end{figure}

In Figure~\ref{low-band-ex-residues}, we make a similar comparison of the low-level residuals for the extended ground plane.  For this ground plane, the worst-case residuals in the bins around Galactic Center transit (LST 16-20~hr) have decreased to $\sim220$~mK compared to $\sim560$~mK for the 10~m~$\times$~10~m case. And during the low-foreground times, the residuals have reduced to $\sim45$~mK compared to $\sim130$~mK from the 10~m~$\times$~10~m case. We see similar trends in the residual levels as the Galactic center moves in and out of the beam, with the most structure when the Galactic center is nearly overhead and the worst agreement between simulation and observation when it is around 45$^\circ$ elevation (LST 14-16~hr and 20-22~hr).  In the extended ground plane case, overall, we see better agreement between data and simulation. Not only are the residuals lower, but the model seems to do a better job matching the data. This is possibly because the soil electrical properties (which are difficult to know exactly) are less of an influence here. At this level, however, we see that the simulated spectra fail to capture all of the spectral features that are evident in the observation residuals. For the 24h average, even after adding the thermal noise of the data (which is $\approx$ 75~mK) in quadrature to the simulated residuals, we still find 14$\%$ unaccountable difference between the residuals. The data residuals tend to have more structure around 65~MHz and 90~MHz compared to the simulations. In the average residual plot we see a larger RMS of the residuals from the data than from the simulated spectra.

\subsubsection{Beam Correction}

When analyzing EDGES data for 21~cm model constraints, we typically apply a beam correction factor to the observed spectra to compensate for the chromatic beam effects.  We follow the procedure described in \citet{Mozdzen_2019} to calculate a frequency-dependent multiplicative factor that adjusts the measured spectrum to yield an estimate of what would have been observed if the beam were perfectly achromatic. In this paper we add to the comparisons the FoMs after correcting for beam chromaticity in both the simulations and the data. To calculate the beam correction factor, we use our same sky model and the simulated beam patterns found by the FEKO solver.  Applying this correction and repeating the FoM calculation serves as a useful additional verification of the performance of the modeled beam patterns.  If we have perfect sky and beam models, the beam correction should return the measured spectrum to its intrinsic smooth spectral shape, reducing the FoM.

In Table~\ref{tabletwo} we  show the FoM values for beam-corrected observations.  As a reference check, we apply the beam correction first to the simulated observations.  For the FEKO simulated observations, we expect that the beam correction should remove all chromatic structure introduced by the beam since the factor is derived from the same underlying model as the simulated observations.  Table~\ref{tabletwo} shows that this is indeed the case.  Applying the beam correction to the actual data, we find that it reduces the FoM for the 10~m x 10~m ground plane from 280~mK to 187~mK, an improvement of 33\%.  This provides further evidence that the simulated beam pattern reproduces some of the structures in the beam of the actual deployed antenna.  For the extended ground plane, which already has a comparatively low FoM of 130~mK, the improvement from beam correction is 8\%.  Our simple sky model may be partly limiting the performance of the beam correction factor.

\subsubsection{Sky Model with Absorption Feature}


While it is not our goal in this work to fully forward model all aspects of the observations, here we do a simple check of including the absorption feature we reported in \citet{bowman2018absorption} in our sky model.  We model the absorption feature as a flattened Gaussian using the equation for $T_{21}(\nu)$ from the Methods section of \citet{bowman2018absorption}. We use the parameter values of $\nu_0 = 78$~MHz, $\tau = 7$, $w = 18$~MHz, and $A = 500$~mK and add the spectral shape as a global contribution to our sky model.  We include these simulated spectra in Figures~\ref{low-band-residues} and~\ref{low-band-ex-residues}. 

For the 10~m~x 10~m ground plane, the residuals after foreground model subtraction are sufficiently large that the addition of the absorption feature has only a minor effect.  In the case of the extended ground plane, the addition of the absorption feature leads to residual spectral structure  that better tracks the data in several LST bins.  In general it seems to help more at 65~MHz than at 90~MHz.  Including the absorption feature improves the agreement of the LST-averaged FoM between the simulations and data. The difference reduces to 15$\%$ compared to 37\% when not including the feature.  If we add to the simulations thermal noise of similar amplitude as in the data, the agreement further improves to 3\% (a difference of only 3.8~mK).  We note that in \citealt{bowman2018absorption}, we also recovered the absorption feature in fits to data averaged over 6~to 18~hr GHA with no beam corrections applied. 

\subsubsection{Confirmation of Soil Properties}

\begin{figure}[h]
\centering
\includegraphics[width=\textwidth]{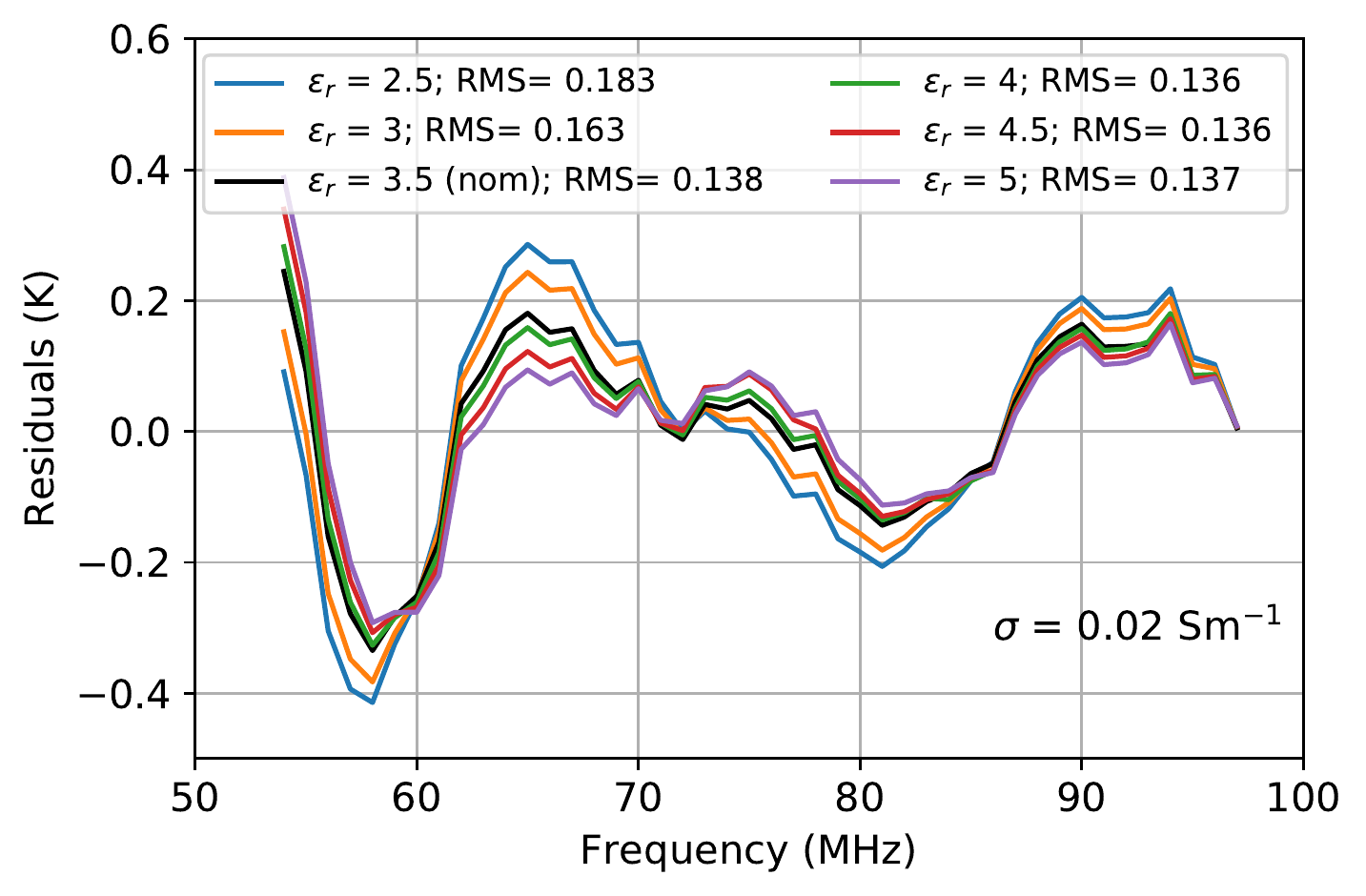}
\includegraphics[width=\textwidth]{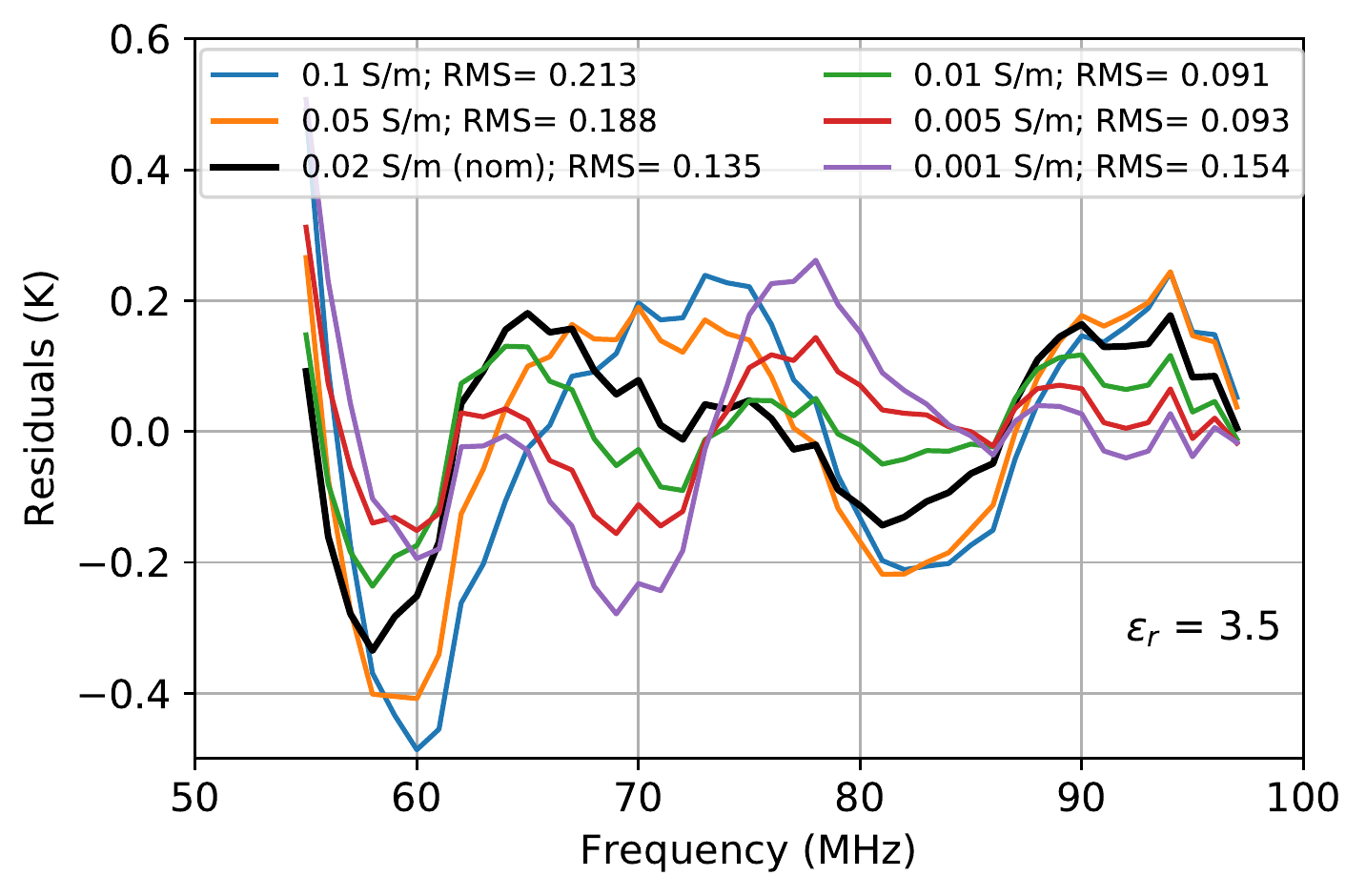}
\caption{Differences between residuals of the simulated and observed data as a function of frequency. The different spectra are generated by varying the soil's relative permittivity (top) and conductivity (bottom). The residuals of the data and simulated spectra are obtained by fitting a five-term Linlog foreground model. The data and beam solutions used in these plots are from the 10~m~$\times$~10~m ground plane system.}
\label{soil-prop}
\end{figure}

All simulations of the antenna that included real soil beneath the ground plane had the soil modeled as a lossy dielectric medium with a relative permittivity ($\epsilon_r$) of 3.5 and conductivity ($\sigma$) of 0.02~Sm$^{-1}$. These are the values for MRO soil reported by \citet{Sutinjo_2015}. Here we test these electrical properties of the soil by comparing simulated spectra generated with different soil parameter values against the observations.  For this comparison, we return to the FEKO simulation and the observations with the 10~m~$\times$~10~m ground plane because the beam patterns for the smaller ground plane are more sensitive to the soil properties than for the extended ground plane.


We perform two grid searches of the soil parameters.  We first vary the relative permittivity from 2.5~to 5~in steps of 0.5, keeping the conductivity constant at 0.02 Sm$^{-1}$, and generate simulated spectra for each of the five different beam solutions.  Following the usual procedure, these spectra are binned in LST and a five-term Linlog model is fitted and removed to obtain the residuals. The residuals of the observations and simulated spectra are averaged over 24 hours. The difference between the residuals from the data and simulated observations is plotted in Figure~\ref{soil-prop}. The RMS of these second-order residuals are indicated in the top panel of the figure to show which value of the soil properties captures the structure in the data residuals. We next repeat the process varying the conductivity from 0.001 Sm$^{-1}$ to 0.1 Sm$^{-1}$ and keeping the permittivity constant at 3.5. The results of this grid search are shown in the bottom panel of Figure~\ref{soil-prop}. In the Figure we see that the simulations are sensitive to the values of the soil parameters. In the top panel, we see that the beam model is not very sensitive to the relative permittivity. Still, $\epsilon_r=3.5$, produces one of the best agreements and lowest RMS across frequency. For conductivity, the curves between 0.005 and 0.02~S/m show the best agreement across frequency. Although 0.005 Sm$^{-1}$ produces a lower RMS, the residuals at lower frequencies are larger. We conclude that the soil properties from \citet{Sutinjo_2015} are representative of the conditions seen by EDGES.

\section{Discussion}

The Cosmic Dawn era is a treasure trove of information of the first stars and early structures in the universe. The redshifted 21~cm line of neutral hydrogen is the most direct probe of this period, but experiments face many challenges.  One of the main challenges for global 21~cm experiments is separating the signal from systematics, particularly those caused by the chromaticity of the antenna beam pattern mixing angular sky structure into the observed spectrum. In this paper, we modeled the EDGES low-band antenna beam using different EM solvers and compared their results. The complete EDGES antenna was simulated over an ideal infinte PEC ground and its response was analyzed using the MoM, FEM, and FDTD numerical techniques.  We found an agreement of within 25$\%$ between four out of the six solutions we computed, which provides reasonable confidence that the calculated beam is not sensitive to the solver technique used.  We then simulated the beams for realistic models of the antenna with the real soil properties and the actual finite ground planes used by the instruments.  We tested three independent MoM solutions and found an agreement to within 15$\%$ for FEKO and CST-I. 



Finite ground planes above realistic soil introduce more beam chromaticity than the ideal infinite PEC ground. In the case of the antenna model with $10$~m $\times$ $10$~m ground plane, the chromaticity is roughly 3.5~times higher than the ideal ground plane as quantified by the FoM of the simulated spectra.  Noting that the residuals from the 10~m~$\times$~10~m ground plane are relatively large compared to the predicted amplitude of the cosmological signal, we showed that larger ground plane sizes can reduce the FoM. Simulations of the extended ground plane used by EDGES predict an LST-averaged FoM of 80~mK, which is only $\approx15$~mK larger than the value for an infinite PEC ground plane and smaller than the amplitude predicted by many models of the 21~cm signature.

To determine if our quantitative analysis of the beam chromaticity captures what we see in our data, we carried out several comparisons. The simulated spectra and sky observations were processed through equivalent pipelines and compared. 
Simulated driftscans agree with observations within 4$\%$ for both ground planes used by EDGES. Additional comparisons were also carried out by fitting the same foreground models to the data and simulated spectra and inspecting the residuals. The LST-averaged FoM from the simulated spectra for the 10~m~$\times$~10~m ground plane is 240~mK, compared to 280~mK from the data. In the case of the extended ground plane, the LST-averaged FoM of the simulated spectra is $\approx$82~mK and increases to $\approx$111~mK after including the absorption feature reported in \citealt{bowman2018absorption} in the sky model.  This addition improves the agreement with the FoM of 130~mK from the data. We also calculated the FoM for the data and simulated spectra after correcting for the beam chromaticity factor. The FoM of the beam corrected spectra showed improvement, providing additional verification of the simulated antenna beam models. Lastly, we looked at the effects of the soil electrical properties on the chromaticity and, by comparing our simulations to measured data, found best-fit soil properties consistent with those reported in \citep{Sutinjo_2015}.


\acknowledgments

This work was supported by the NSF through research awards for EDGES (AST-1207761, AST-1609450, AST-1813850, and AST-1908933). N.M. was supported by the Future Investigators in NASA Earth and Space Science and Technology (FINESST) cooperative agreement 80NSSC19K1413.  EDGES is located at the Murchison Radio-astronomy Observatory. We acknowledge the Wajarri Yamatji people as the traditional owners of the Observatory site. We thank CSIRO for providing site infrastructure and support.

\newpage

\bibliography{Main}{}
\bibliographystyle{aasjournal}

\end{document}